\documentclass[fleqn,usenatbib]{mnras}

\usepackage{enumerate}
\usepackage{xcolor}
\usepackage{amsmath}

\usepackage{newtxtext,newtxmath}

\usepackage[T1]{fontenc}

\DeclareRobustCommand{\VAN}[3]{#2}
\let\VANthebibliography\thebibliography
\def\thebibliography{\DeclareRobustCommand{\VAN}[3]{##3}\VANthebibliography}

\usepackage{graphicx}	
\usepackage{amsmath}	

\title[Short title, max. 45 characters]{Segmentation of spectroscopic images of the low solar atmosphere by the Self Organizing Map technique}

\author[F. Schilliro et al.]{
F. Schilliro,$^{1}$\thanks{E-mail: francesco.schilliro@inaf.it}
P. Romano,$^{1}$
\\
$^{1}$Osservatorio Astrofisico di Catania, INAF, via S. Sofia n.78, 95123 Catania, Italy\\
}

\date{Accepted XXX. Received YYY; in original form ZZZ}

\pubyear{2020}

\begin{document}
\label{firstpage}
\pagerange{\pageref{firstpage}--\pageref{lastpage}}
\maketitle

\begin{abstract}
We describe the application of Semantic Segmentation by using the Self Organizing Map technique to an high spatial and spectral resolution dataset acquired along the H$\alpha$ line at 656.28 nm by the Interferometric Bi-dimensional Spectrometer installed at the focus plane of the Dunn Solar Telescope. This machine learning approach allowed us to identify several features corresponding to the main structures of the solar photosphere and chromosphere. The obtained results show the capability and flexibility of this method to identifying and analyzing the fine structures which characterize the solar activity in the low atmosphere. This is a first successful application of the SOM technique to astrophysical data sets.
\end{abstract}




\section{Introduction}

With the rise in big data, machine learning has become particularly important for solving data science problems in different areas (Finance, Image processing, Computational biology, Automotive, Aerospace, Natural language processing) \citep{Agg01}. Machine learning algorithms and techniques allow scientists to automatize computer learning in order to make classifications, decisions and predictions without the continuous interaction with systems under analysis.  
In general machine learning, as well semantic segmentation in particular, use two types of approaches: supervised learning, almost focusing on the 'deep learning' methodologies, e.g. based on Convolutional Networks \citep{Hau1, She01}, and unsupervised ones. The former algorithms train a chain of neural networks by processing a known input and correlated output data, so that it is possible to predict future outputs by learning from 'experiences'. Instead, less used unsupervised learning method \citep{Bar01, Zel01}, find hidden patterns or intrinsic structures inside the input data, allowing data classification or segmentation in a non 'a priori' organized fashion. 

The application of machine learning techniques has also a great potentiality in the space weather field in general, and for solar images in particular, with the opportunity to identify new indicators for the forecast of eruptive events occurring on the Sun, such as flares and Coronal Mass Ejections (see \citet{Benz17}), and producing effects in the near-Earth space environment, as well as at the Earth surface. 

In fact, the scientific community has developed a growing interest for the automatic identification of the signatures characterizing the sources of disturbances whose radiative, field, and particle energy directly impact the Earth \citep[e.g.,][]{Bar16, Kim19, Korsos19, Flo18, Kon18, Fal19}. In particular, the characterization of solar active regions allows to identify the topological configuration which are suitable for the occurrence of sudden, rapid, and intense solar eruptions \citep[e.g.,][]{Rom15, Rom18, Rom19}.

In this paper, we consider for the first time the application of the Semantic Segmentation by using 'Self Organizing Maps' (SOM) technique \citep{Koh01}, to high-resolution spectroscopic observations of the low solar atmosphere. We used a data set taken during an observing campaign carried out on May 2015 at the NSO/Dunn Solar Telescope (DST). The data set showing a sunspot and its main structures visible at photospheric and chromospheric level is described in the next Section. We describe the technique applied to get a skeletrization of the main structures in Section 3. Section 4 reports the main results and the advantages of the application of this new technique to solar images. The last Section is dedicated to draw our conclusions.




\section{Data}

We used data acquired by the Interferometirc Bidimensional Spectroscopic Instrument \citep[IBIS;][]{Cav06} which operated at the NSO/DST. The data set consists of high-resolution data, with a pixel scale of 0\farcs095, taken on 2015, May 18, in 17 spectral points along the H$\alpha$ 656.28 nm line \citep[see][for further details]{Rom17}. 

We selected the scan of the H$\alpha$ line taken during the best seeing conditions of that day of observation. 

The original field-of-view (FOV) of IBIS data was 500$\times$1000 pixels, but we considered in our analysis only a sub-FOV of 350$\times$700 pixels centered on the preceding sunspot of the AR NOAA 12348. In this way it was possible to analyze the results obtained by the application of our method of analysis to the main structures visible in a sunspot and in the surrounding environment \citep{Rom20}.

We also used broadband images at 633.3 nm acquired simultaneously with the spectral frames, imaging the same FOV with the same exposure time, in order to restore the spectral images using the Multi-frame Blind Deconvolution \citep[MFBD;][]{Lof02} technique. The MFBD allowed us to reduce the seeing degradation, achiving a spatial resolution of about 0\farcs3 at 656.28 nm. 

\begin{figure}
\begin{center}
\includegraphics[trim=0 160 170 80, clip, scale=0.22]{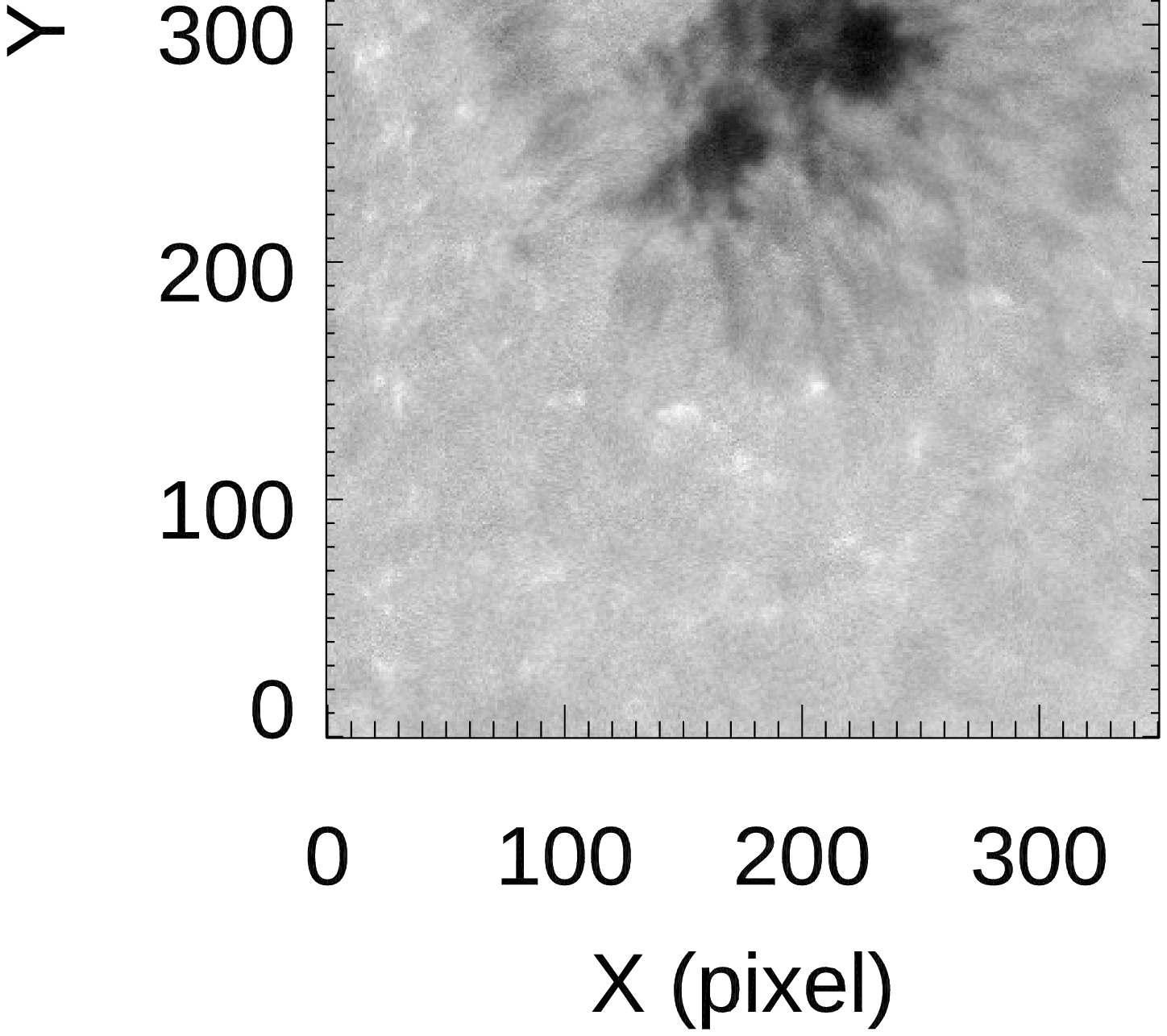}
\includegraphics[trim=108 160 170 80, clip, scale=0.22]{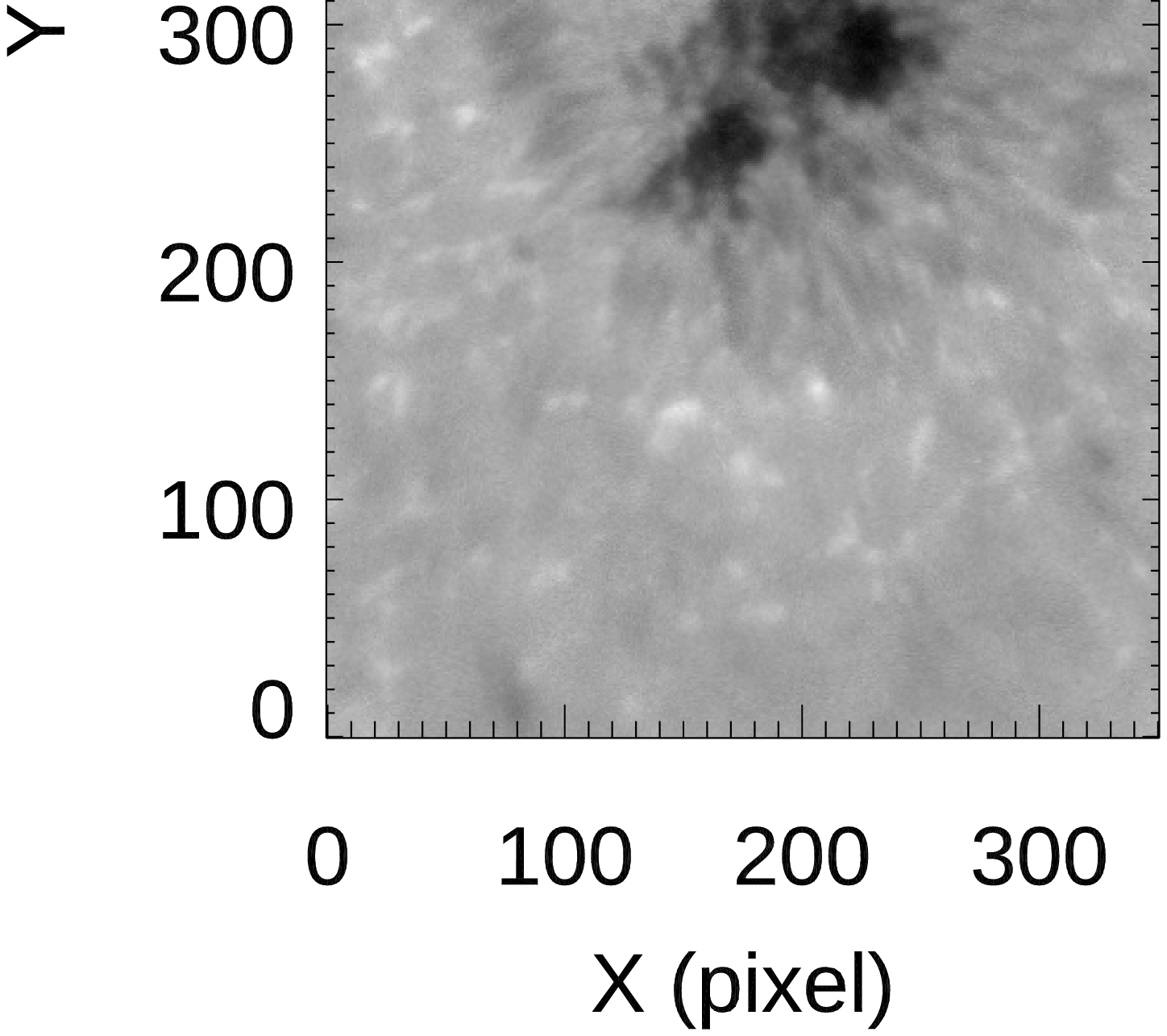}
\includegraphics[trim=108 160 170 80, clip, scale=0.22]{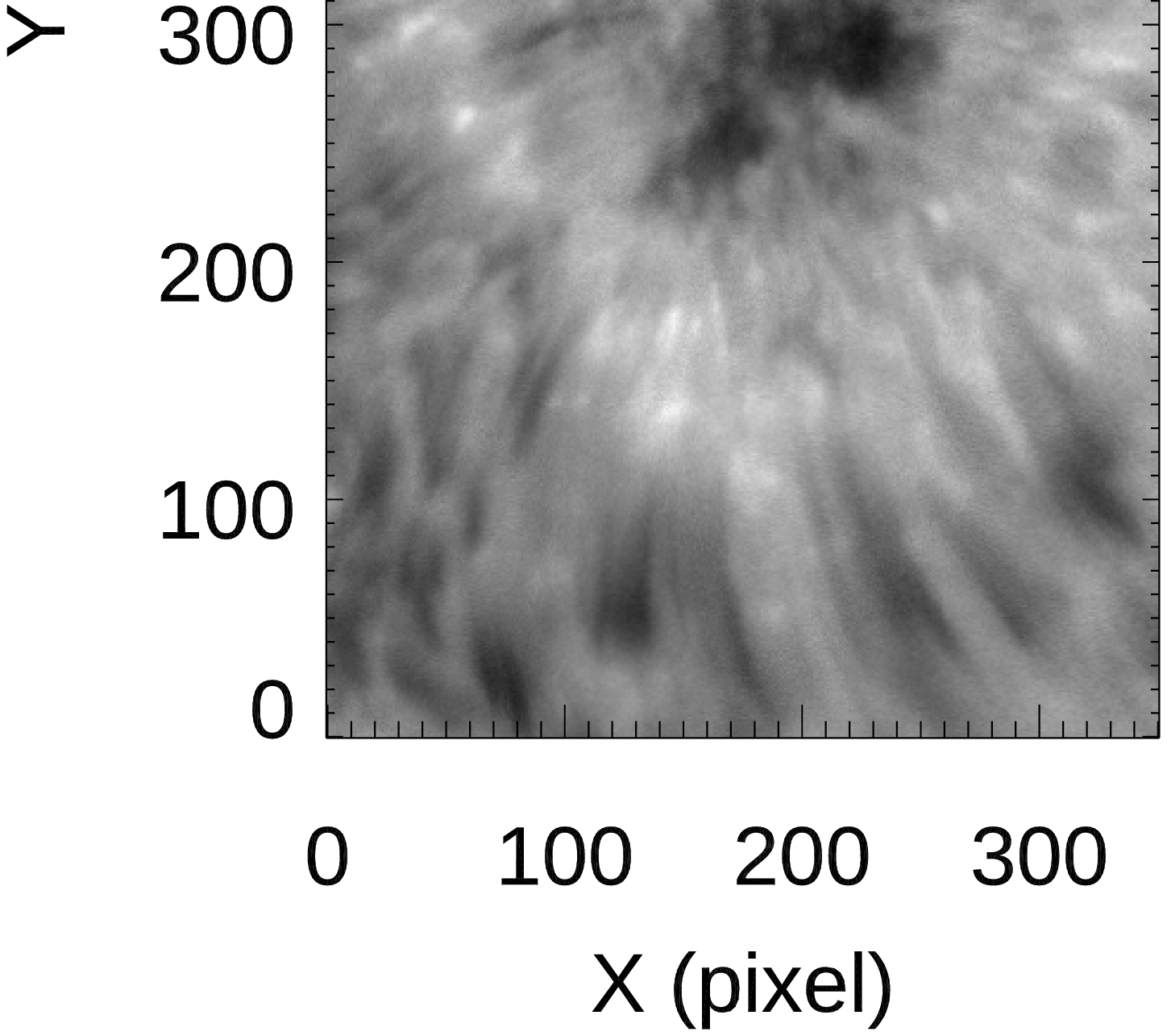}\\
\includegraphics[trim=0 160 170 80, clip, scale=0.22]{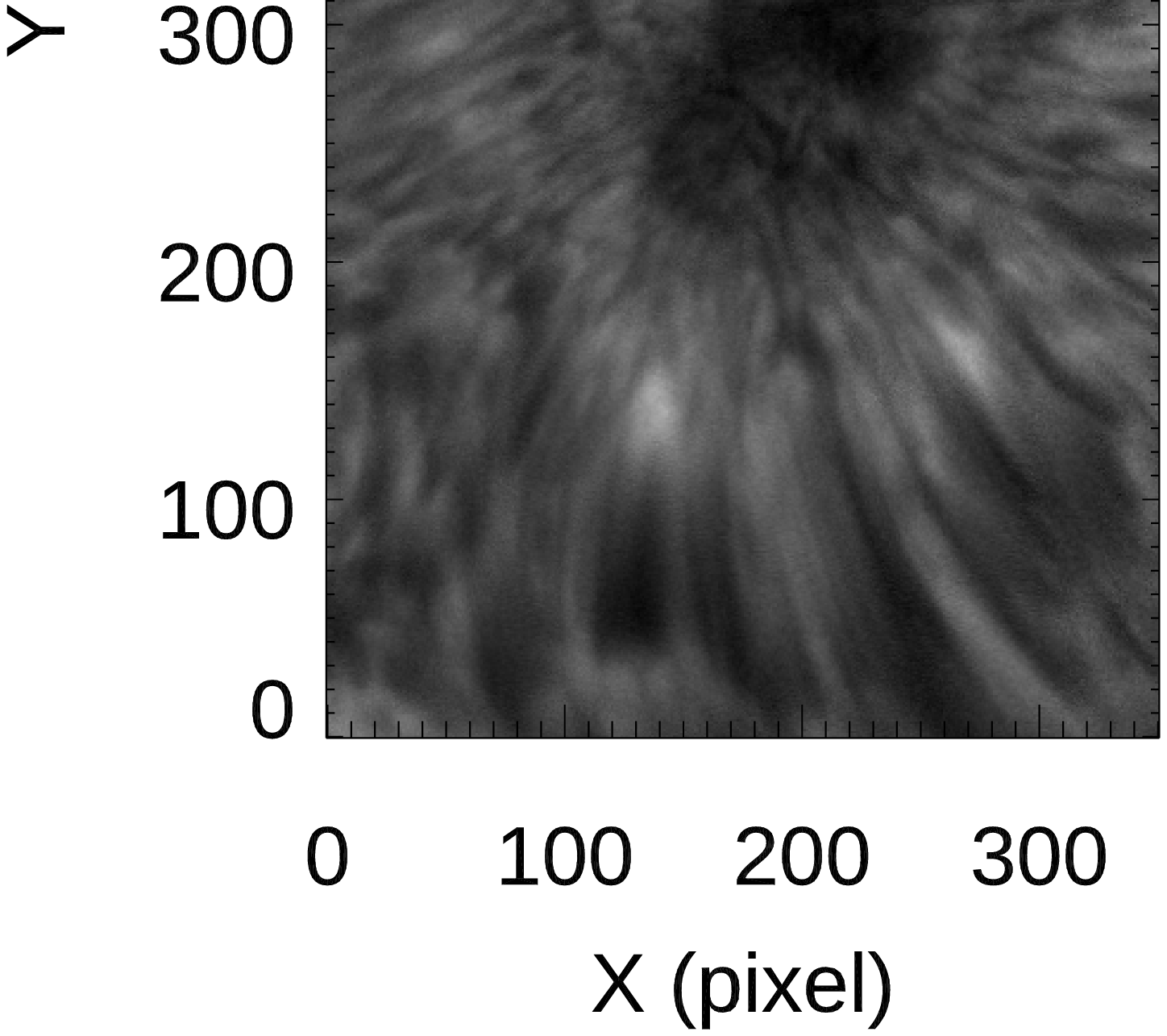}
\includegraphics[trim=108 160 170 80, clip, scale=0.22]{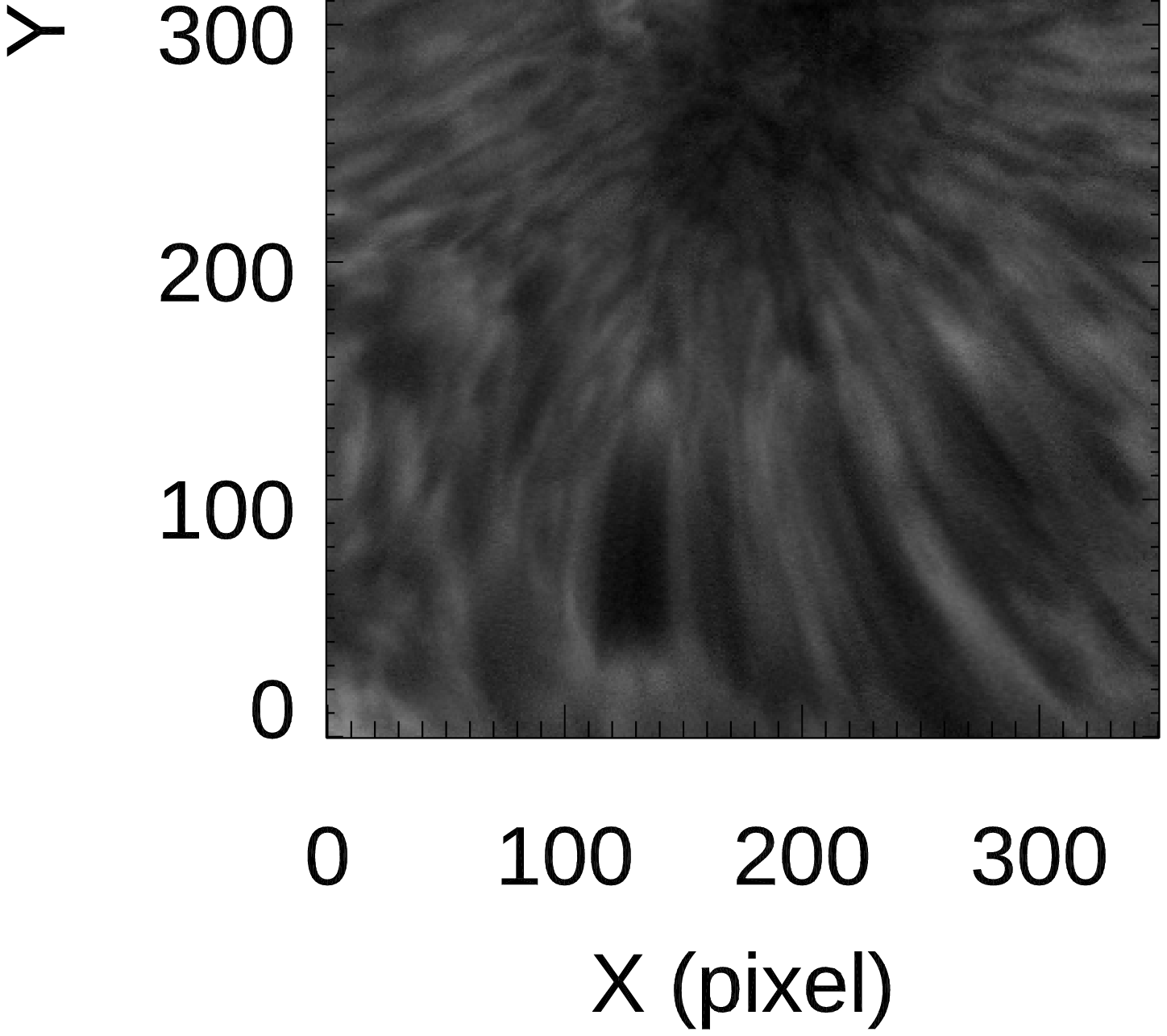}
\includegraphics[trim=108 160 170 80, clip, scale=0.22]{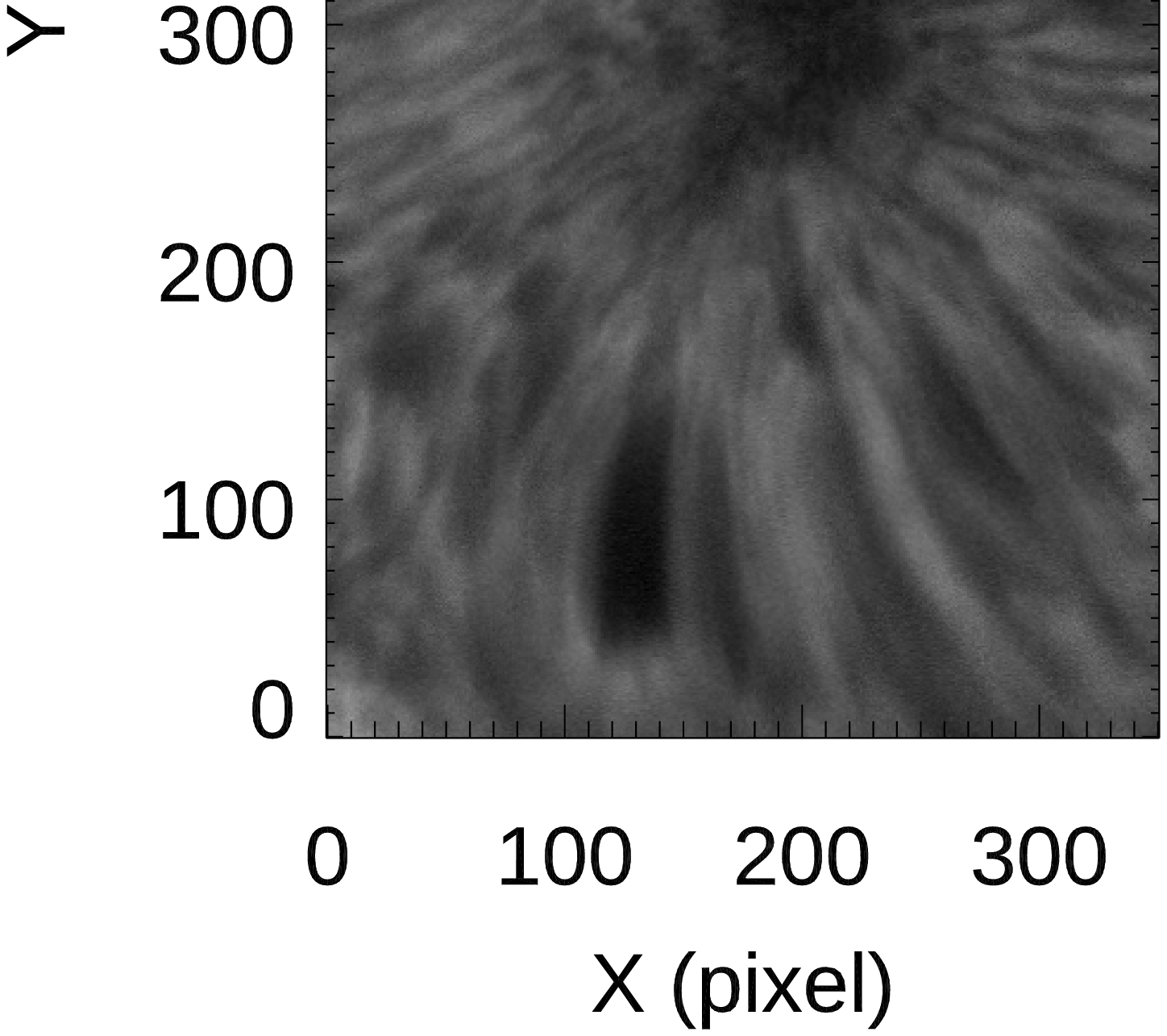}\\
\includegraphics[trim=0 65 170 80, clip, scale=0.22]{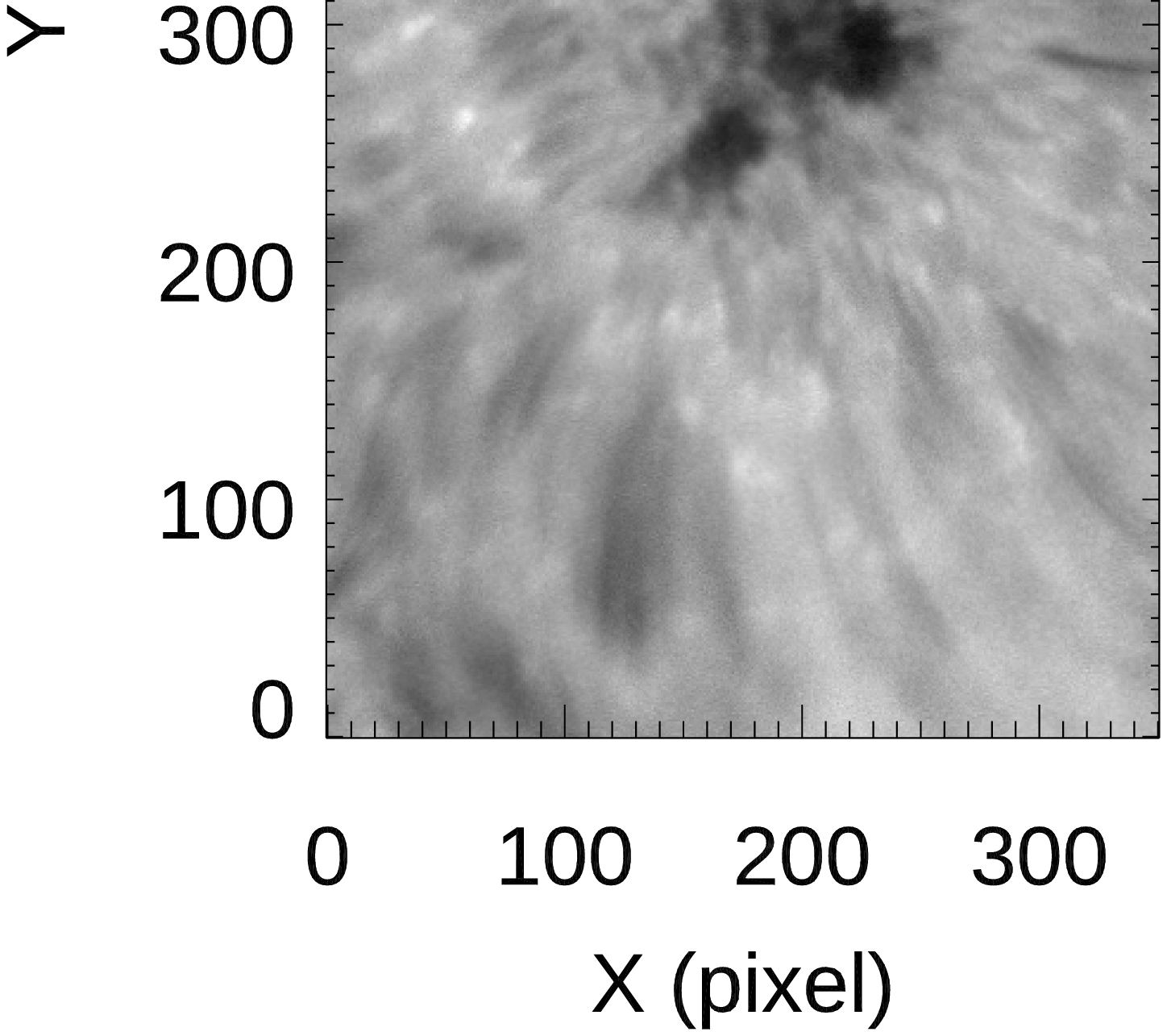}
\includegraphics[trim=108 65 170 80, clip, scale=0.22]{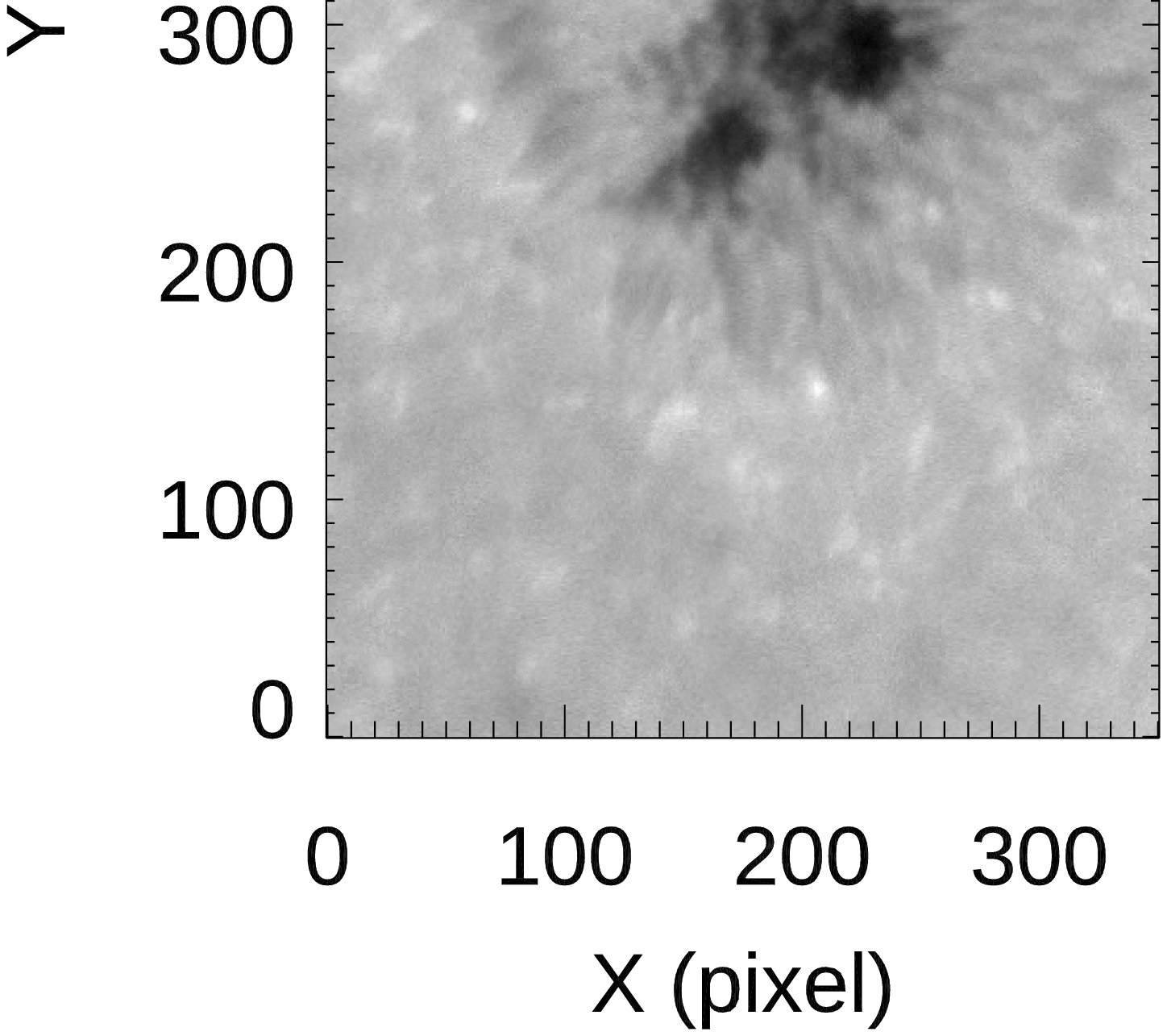}
\includegraphics[trim=108 65 170 80, clip, scale=0.22]{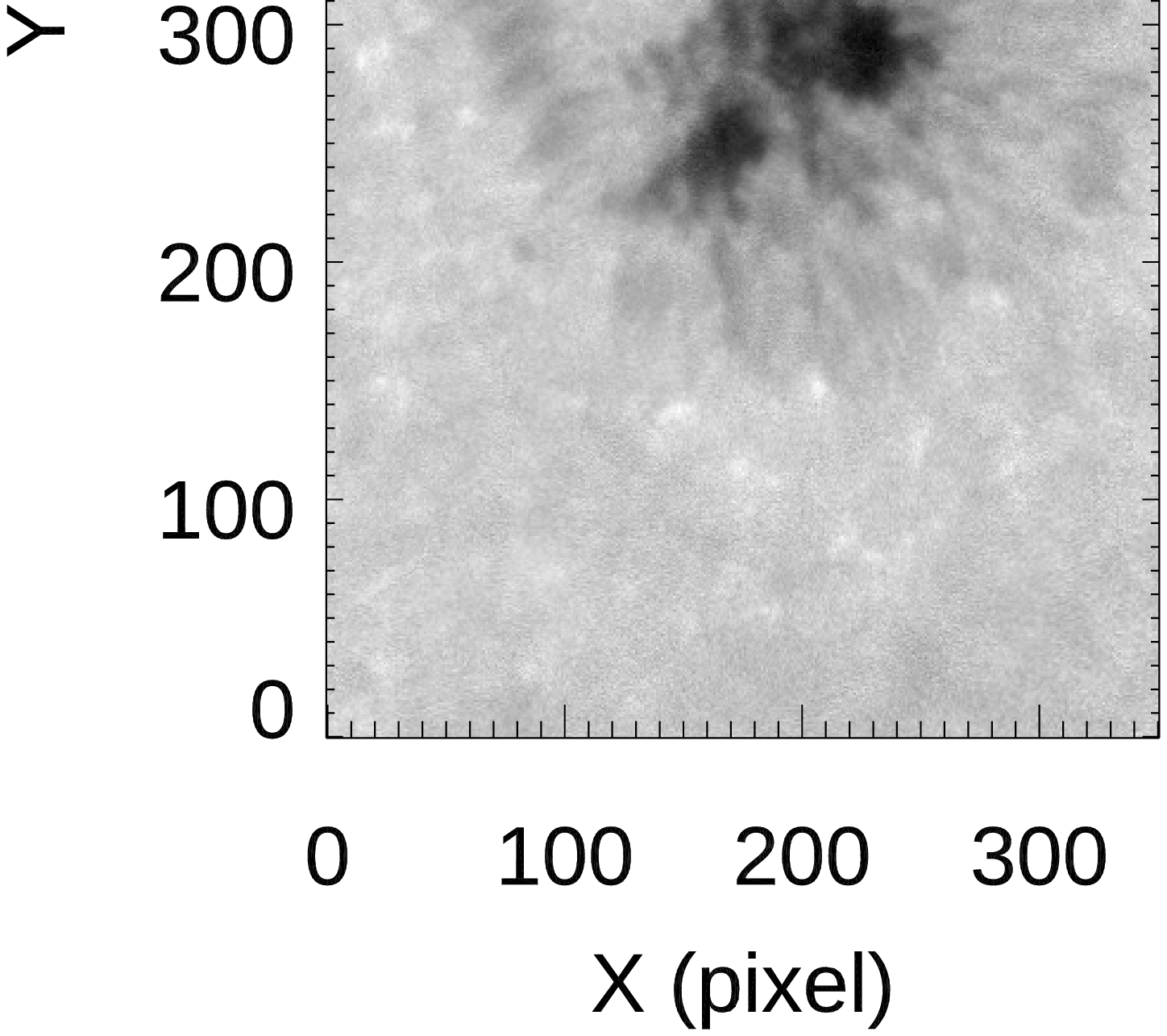}\\

\caption{Sample of the spectral images acquired along the H$\alpha$ line by IBIS and used in our analysis. All the images have been acquired on May 18 at 14:42 UT. \label{fig1}} 
\end{center}
\end{figure}

In Figure \ref{fig1} we report some of the spectral images used for our analysis. We can see the different solar structures visible in our FOV varying the spectral points along the H$\alpha$ line. In the spectral points near the continuum (top left and bottom right panels of Figure \ref{fig1}) and in the wings of the spectral line (top right and bottom left panels of Figure \ref{fig1}) we are able to identify the main fine structures of the sunspot and the surrounding environment typical of the photospheric level, such as the umbra, the penumbral filaments, a light bridge, the granular and intergranular pattern \citep[see][for an overview of these structures]{Sol03}. Near the center of the H$\alpha$ line (middle row panels of Figure \ref{fig1}) we clearly see the chromospheric structures, such as the superpenumbra, some facular regions in the top right corner of the FOV, and a filament portion in the top left corner of the FOV \citep[see][]{Lek03}.

In order to better describe the potentiality of the SOM technique in the segmentation of our spectral dataset, we also considered some physical parameters which can be deduced by the analysis of the H$\alpha$ line profile. We reconstructed the profile of the H$\alpha$ line in each spatial pixel by fitting the corresponding signals obtained in the monochromatic images with a linear background and a Gaussian shaped line. Using this reconstructed profile we performed an estimation of the line depth (LD) by:
\begin{equation}
 \frac{(F_c-F_o)}{<I>},
\end{equation}
where $F_{c}$ and $F_{o}$ are the maximum and the minimum intensities of the line profile, taken in the continuum and in the centroid of the gaussian fit, respectively, while $<I>$ is the average intensity of the quiet Sun, measured in a box of 10 $\times$ 10 pixels taken in the southern portion of the FOV, outside of the sunspot where few chromospheric structures were visible. Using the full width at halpha maximum (FWHM) of the gaussian fit we estimated the spectral line width (LW).
We also measured the value of velocity along the line of sight (LOS) in photosphere from the Doppler Shift (DS) of the centroid of the reconstructed line profiles in each spatial point with respect to the reference wavelength (656.28 nm).

\section{The Self Organizing Map technique}
Self-Organizing Maps (SOM) is an important class of neural networks capable of unsupervised autonomous learning from data, adapting according to rules of synaptic plasticity in response to external stimuli.
SOM algorithms are inspired by neuro-biological studies and indicate that different sensory inputs (motor, visual, auditory, etc.) are mapped onto corresponding areas of the cerebral cortex in an orderly fashion.
This form of map, known as a topographic map, has two important properties:
\begin{enumerate}
  \item At each stage of representation, or processing, each piece of incoming information is kept in its proper context/neighbourhood.
  \item Neurons dealing with closely related pieces of information are kept close together so that they can interact via short synaptic connections.
\end{enumerate}

\begin{figure}
\begin{center}
\includegraphics[trim=0 0 50 0, clip, scale=0.45]{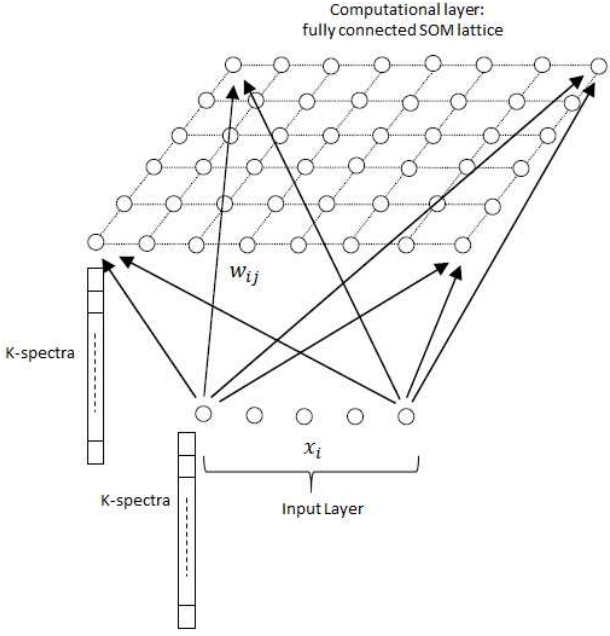}

\caption{Neural lattice of a Kohonen Network: the $N_1\times N_2$ fully connected lattice compute K-spectra vector input layers as vector of spectral values associated to each pixel of image under processing.\label{fig2}} 
\end{center}
\end{figure}

A particular kind of SOM known as a Kohonen Network \citep[see][for an overview]{Koh01}, belongs to the class of vector-coding algorithms; they offer a mapping topological which places a fixed number of vectors (code words) in an input space oversized, thus allowing data compression, data clustering or data classification. This ability to extract information from the data and understand how these group spontaneously into cluster, is allowed by a neural structure with a single computational layer arranged in rows and columns, in which each neuron is fully connected to other and to all the source nodes in the input layer.
The self-organization algorithm is composed in general of four major process that involves input layer, as source of data to be processed and neural fully connected lattice, that adapts its structure and morphology in order to replicate the input organization of data according to their neighborhood.  
In our study, this lattice morphology is used to semantic grouping pixel in regions, the features, that are detected according to the distance of the multi-spectral set of input data collected for each pixel and the normalized random initial formatting of the net, as depicted in Figure~\ref{fig2}.

1. Initialization: Input pattern is a normalized data-cube composed by a K-dimensional vector of spectral values for all the N$\times$M pixels, as described in Fig.2. All the K-dimensional connection weights for each neuron and for each input value, \(w_{ji}\), are initialized with normalized random values.

2. Competition: For each input pattern, the neurons compute their respective values of a discriminant function which provides the basis for competition.
For the D=N$\times$M pixel/neurons of  K-dimensional spectral input space, discriminant function is calculated as the spectral Euclidean distance between each input pattern $(x=(x_i:i=1,…,D))$ and each connection weights between the input units i and the neurons j in the computation layer, $(w_{ji}={x_{ji}: j=1…N_1\times N_2 , i=1,…,D})$, were $(L=N_1\times N_2)$ is the neural lattice dimension.

\begin{equation}
 \Delta_i(\bar{x})=\displaystyle\sum_{i=1}^{D} (\bar{x_i}-\bar{w}_{ji})^2
\end{equation}

The neuron $E(x)$ whose weight vector comes closest to the input vector (i.e. is most similar to it) is declared the winner, so the continuous input space can be mapped to the discrete output space of neurons by a simple process of competition between the neurons.

3.	Cooperation: The winning neuron $E(x)$ determines the spatial location of a topological neighbourhood of excited neurons, thereby providing the basis for cooperation among neighbouring neurons, what in neuro-biological studies is the lateral interaction within a set of excited neurons. When one neuron fires, its closest neighbours tend to get excited more than those further away, and this decays with distance according a gaussian law :

\begin{equation}
G_{\delta,E(x)}=\mathrm{e}^\frac{-S^2_{\delta,E(x)}}{2\sigma^2} 
\end{equation}

where

\begin{itemize}
  \item $E(x)$ is the winning neuron according $x$ input vector;
  \item $\delta$ is the general index for neuron of the lattice;
  \item  $\mathrm S_{\delta,E(x)}$ is the lateral distance between $E(x)$ and $\delta$, calculated as the Euclidean distance on each spectral K-dimensional vector;
  \item  $\sigma$ is the size of topological neighbourhood, typically shrinking with time (expressed in number of iterations as well) as the function $\sigma(t)=\mathrm{e}^\frac{-t} {\tau_0}$ , with $\sigma_0$ and $\tau_0$ constant set by the neural network designer and respectively the initial neural proximity parameter and the mean time of activation persistence by proximity.
\end{itemize}
Cooperation process is depicted in Figure 3, where is highlighted the group of involved neurons by application of gaussian topological neighbourhood operator.   

\begin{figure}
\begin{center}
\includegraphics[trim=0 0 0 0, clip, scale=0.45]{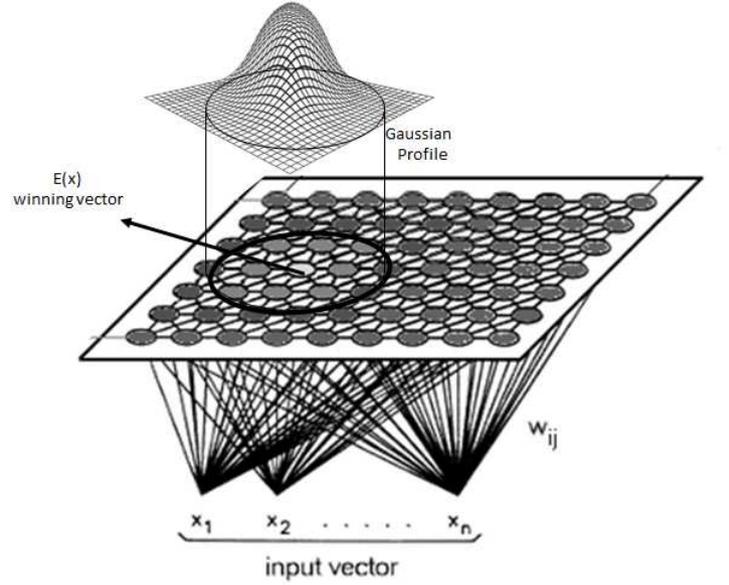}
\caption{Group of cooperating neurons by application of Gaussian topological neighbourhood operator.\label{fig3}} 
\end{center}
\end{figure}

4.	Adaptation

The excited neurons decrease their individual values of the discriminant
function in relation to the input pattern through suitable adjustment of the associated
connection weights, such that the response of the winning neuron to the subsequent
application of a similar input pattern is enhanced.
The adaptation of the SOM designed is the process responsible for the feature map formation, that links the input data to the self-organized neuronal lattice.
The point of the topographic neighbourhood is that not only the winning neuron gets its weights updated, but its neighbours will have their weights updated as well, although by not as much as the winner itself. In practice, the appropriate weight update equation in term of weights updating is

\begin{equation}
\Delta w_{ji}=\eta(t)G_{\delta,E(x)}(t) (\bar{x}-\bar{w}_{ji})
\end{equation}

\begin{equation}
\eta(t)=\eta_0\mathrm{e}^\frac{-t} {\tau_n}
\end{equation}

Upon repeated presentations of the training data, the synaptic-weight vectors $w_j$ tend to follow the distribution of the input vectors because of the neighborhood updating. The algorithm therefore leads to a topological ordering of the feature map in the input space in the sense that neurons that are adjacent in the lattice will tend to have similar synaptic-weight vectors, towards the input vector $x$.

The learning-rate parameter $\eta(t)$ is responsible for speed of adaptation process, should also be time varying, as indicated in equation (4), and is time shrinking according the learning ability $\eta_0$ and its persistence in time $\tau_n$. 

\section{Clustering criterion and SOM  features selection }

The operation of image segmentation is basically the association of each pixel $x_{i}$ of a frame, that represents a K-vector spectral data set, to a unique cluster $C_{l}$ according to a method that in our case is a SOM. A Segmented region associated to each different cluster could be expressed as ‘feature’ $\rho_{k,i}$ as follows:

$$\rho_{k,i}= \left\{
  \begin{array}{lr} 
      1 & \forall x_i \in C_l \\
      0 & \text{otherwise} 
      \end{array}
\right.$$

so that it is possible to define the number of pixel belonging to $C_l$ as  $|C_l|=\displaystyle\sum_{i=1}^{L} \rho_{l_i}$ 
 
In Cluster Analysis two strategies are possible in finding a possible optimal data classification, a supervised and an unsupervised approach. While the first one  is adopted when segmentation of region in features is well specified and tested, the unsupervised one is a method in which the number of features is not previously specified, and it arises from data structure as a results of a data processing that produces different indices of clustering. 

In our case the number of clusters or 'features' in not known 'a priori', because a class is directly connected to the presence of a region of interest that could be in the tile under study or not. Therefore, it could be useful to find inside the data structure, whether an optimal number of possible groups of data separated each others exists , each of which corresponding to different regions composing the image frame.    

For the purpose of discovering the number of independent regions, that becomes  important in the stage of design the SOM network, different clustering indices and parameters were investigated, as the Calinski-Harabasz (CH) index \citep[see][]{Cal01}, the Silhouette index \citep[see][]{Rou1}, and Davies-Bouldin (DB) index \citep[see][]{DB1}. In particular, DB has been already used in literature for solar EUV data clustering \citep[see][]{Cab1}, confirming that is a good and fast algorithm which converges with the analysed data. 
In our case, we found that DB index provided same results in the classification of optimal number of independent regions considering not only the clustering of the N$\times$M$\times$K spectral data-cube, but also the clustering of the N$\times$M$\times$3 data-cube containing the measured parameters of the spectral line (LW, LD and DS). 

\begin{figure}
\begin{center}
\includegraphics[trim=0 0 1450 0, clip, scale=0.18]{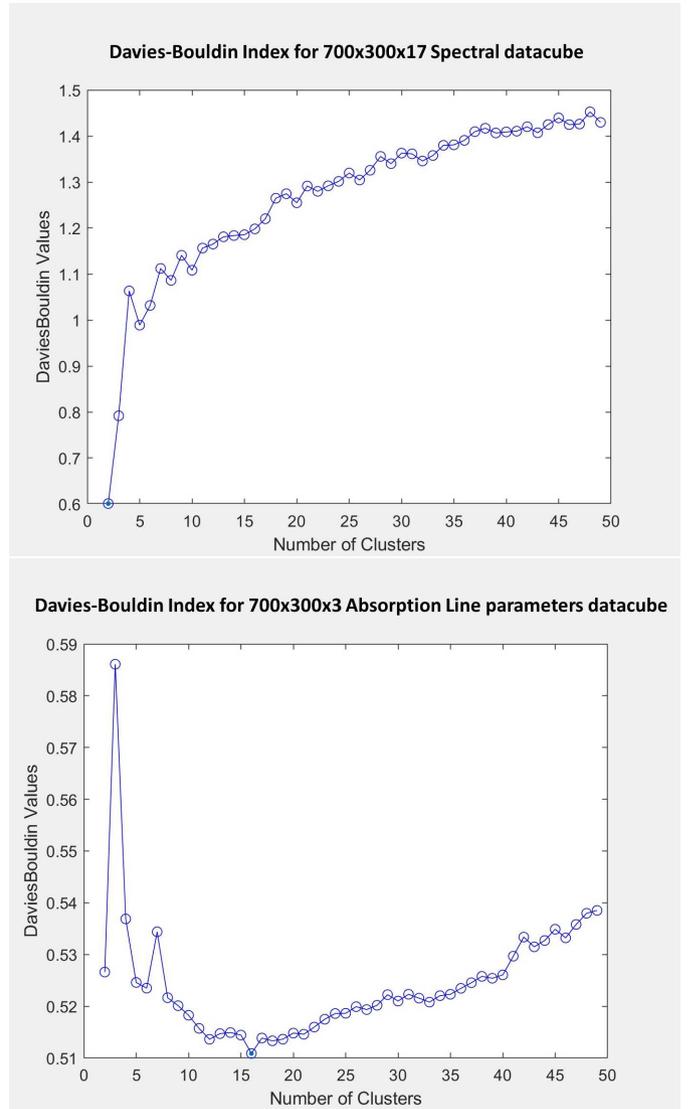}\\
\includegraphics[trim=1400 0 50 0, clip, scale=0.18]{fig4_1.eps}
\caption{Davies-Bouldin index for multi-spectral (up) and parameters (bottom) datacube .\label{fig4}} 
\end{center}
\end{figure}

In Figure \ref{fig4} we report the graphs of the DB index for the spectral data-cube and the parameters data-cube. They show the best result (i.e., a minimum) when the number of clusters is equal to 16, which can be considered the best number of regions for the segmentation and design of the SOM lattice. 
Anyway using a more (less) populated neural lattice with a number of features more (less) than 16, makes effective the semantic segmentation splitting (merging) the different regions in components that keep the same nature.  

The algorithm complexity and relative time of execution is not hardly affected by SOM network complexity, that depends on the lattice structure. The choice of a number of features according to relative minima in the DB index is determined by the required detail level for the segmentation of the data; in our case we considered with particular attention the case of 16 regions designing a 4x4 SOM neural lattice.
The choice of $N_1=4$ and $N_2=4$, complete the overall architecture for the segmentation algorithm, that really needs some trimmer for the values related to $\sigma_0$, $\tau_0$, $\eta_0$ , $\tau_n$. 
This architecture is depicted in Figure \ref{fig5}.

\section{Results}

\begin{figure}
\begin{center}
\includegraphics[trim=0 0 0 0, clip, scale=0.28]{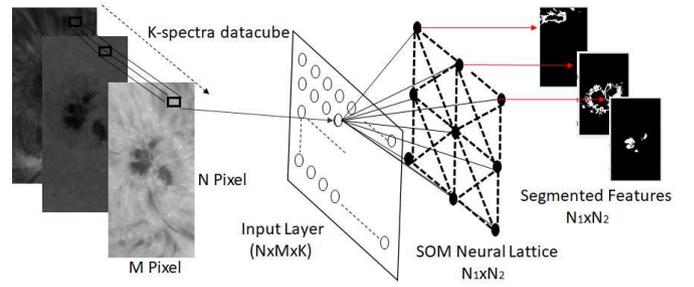}
\caption{Architecture for a Semantic Segmentation using Self Organizing Map with a 4x4 computational neural lattice.\label{fig5}} 
\end{center}
\end{figure}

The algorithm defined in Figure \ref{fig5} runs after a phase of preparation of the input data cube which foresees the adaptation of the image to the appropriate size (in this case 350x700 pixels) and the normalization of the brightness values of each pixel for each wavelength of the data cube.
The SOM network is organized in such a way as to guarantee segmentation in complementary regions, so that the results of the elaboration are two-dimensional maps, which display the features, i.e., semantically separated regions.

\begin{figure}
\begin{center}
\includegraphics[trim=0 0 0 0, clip, scale=0.20]{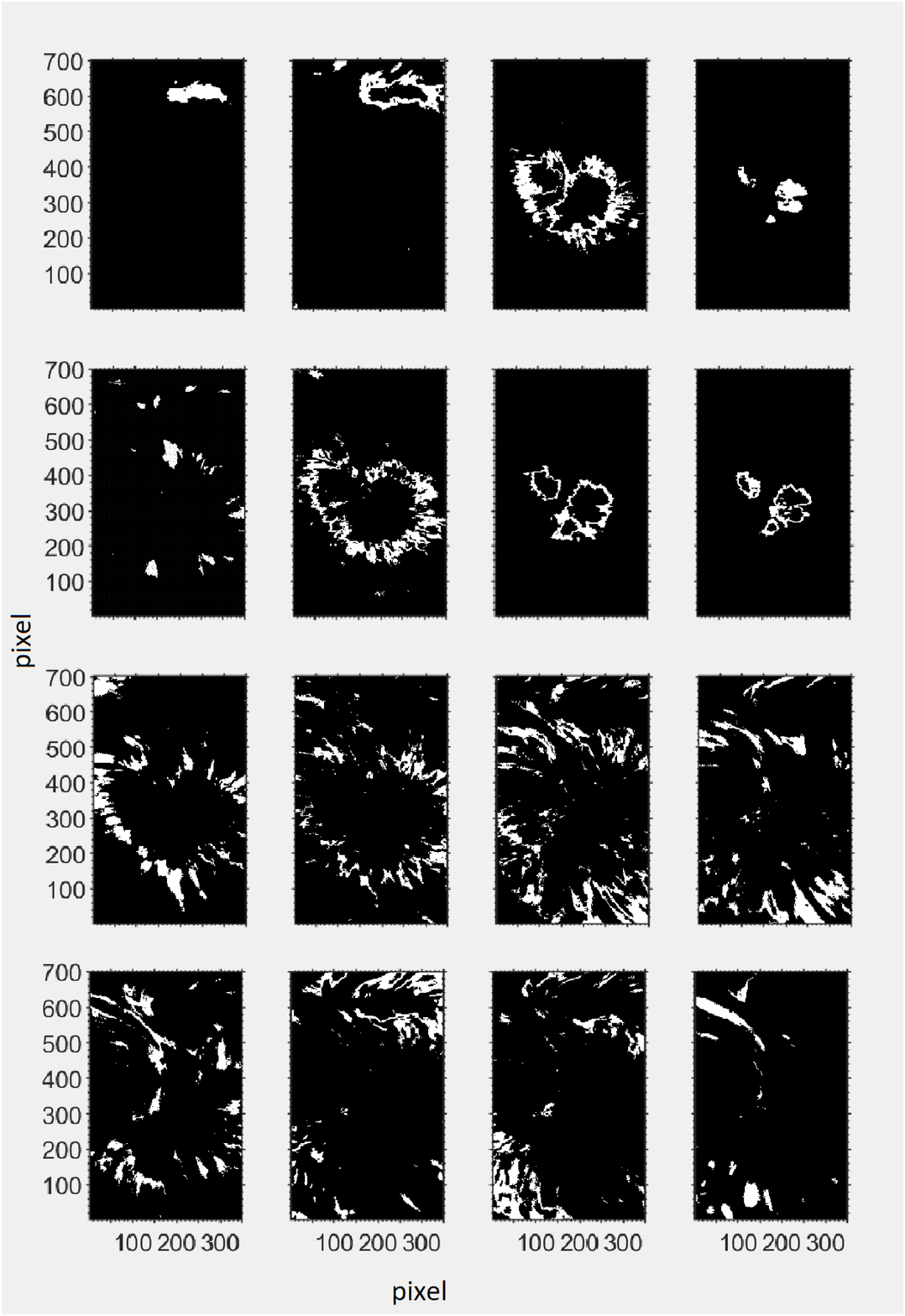}
\caption{Feature lattice elaborated by 4x4 SOM and performing the Semantic Segmentation of 16 different regions.\label{fig6}} 
\end{center}
\end{figure}

\begin{figure}
\begin{center}
\includegraphics[trim=0 0 0 0, clip, scale=0.20]{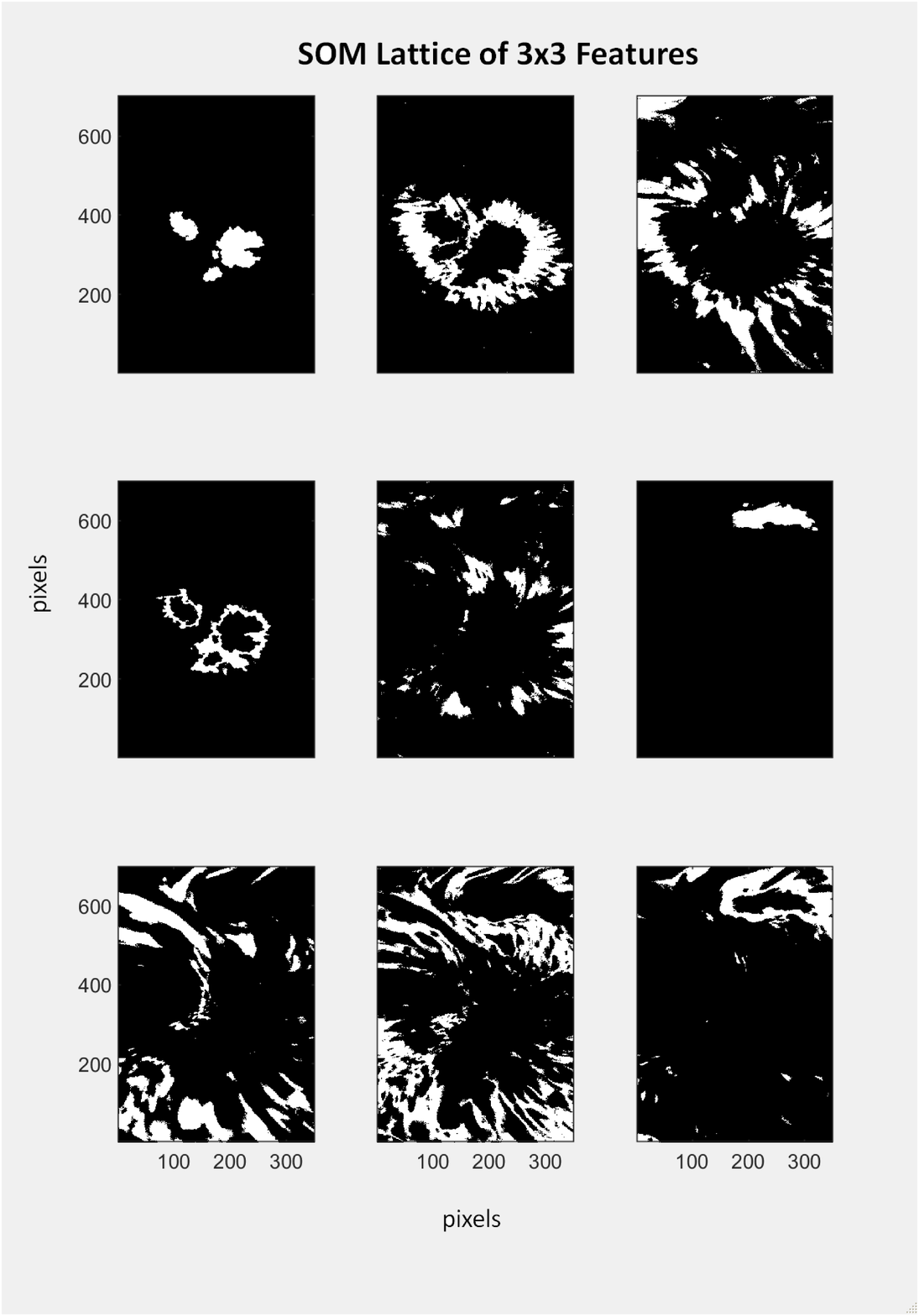}
\caption{Feature lattice elaborated by 3x3 SOM and performing the Semantic Segmentation of 9 different regions.\label{fig6bis}} 
\end{center}
\end{figure}

One of the main advantage of the SOM technique is its flexibility in the choice of the number of features. In fact, according to the needs of the user it is possible to segment the dataset with different level of accuracy. As mentioned above, on the base of the DB index we decided to consider the results obtained for 16 clusters (Figure \ref{fig6}) and to describe in detail the different features found in this particular case. However, a different number of clusters, e.g., 9 (Figure \ref{fig6bis}), results in the merging of some features obtained for a higher value (compare Figures \ref{fig6} and \ref{fig6bis}). 

\begin{figure*}
\begin{center}
\includegraphics[trim=0 0 0 0, clip, scale=0.16]{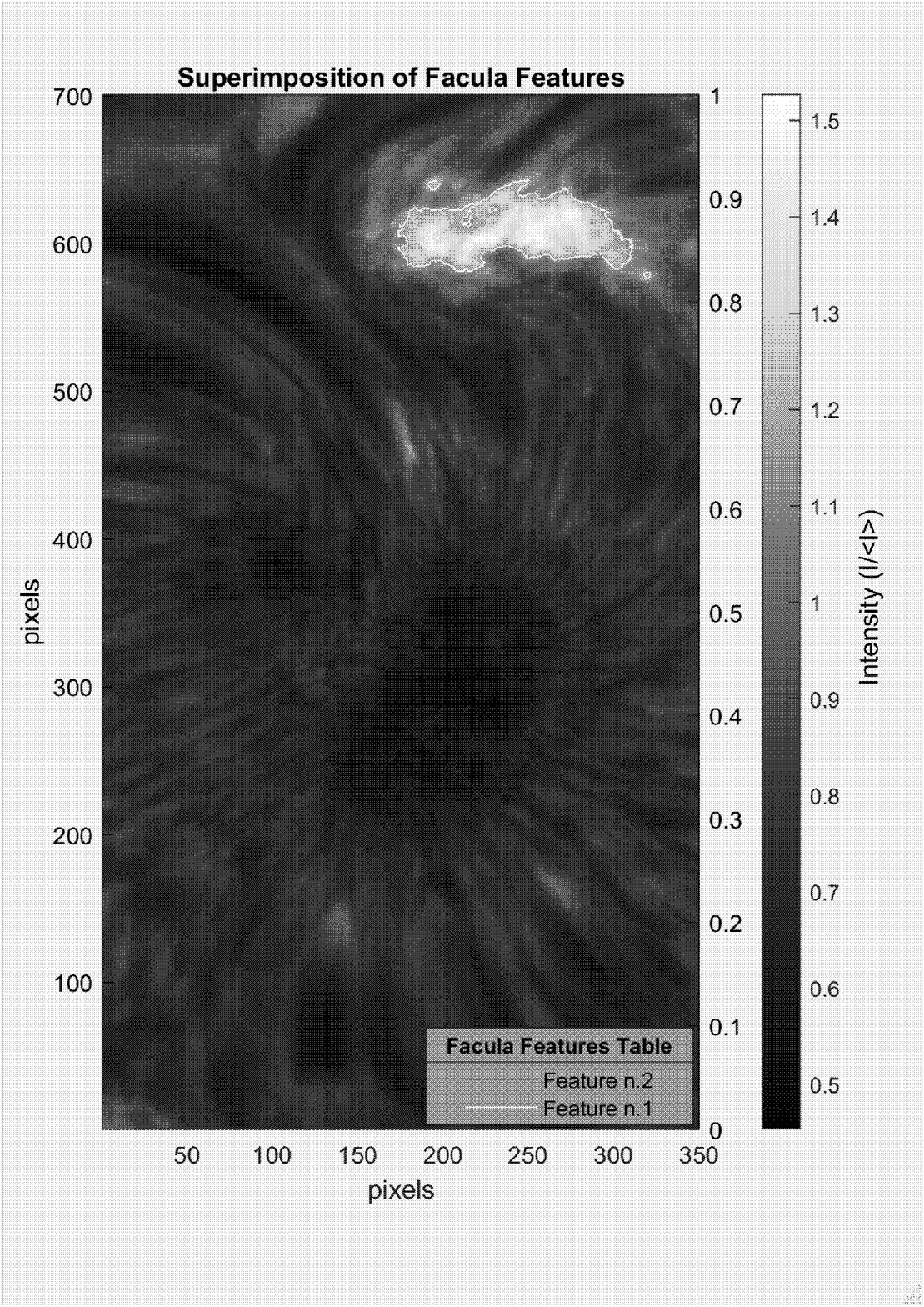}
\includegraphics[trim=0 0 0 0, clip, scale=0.16]{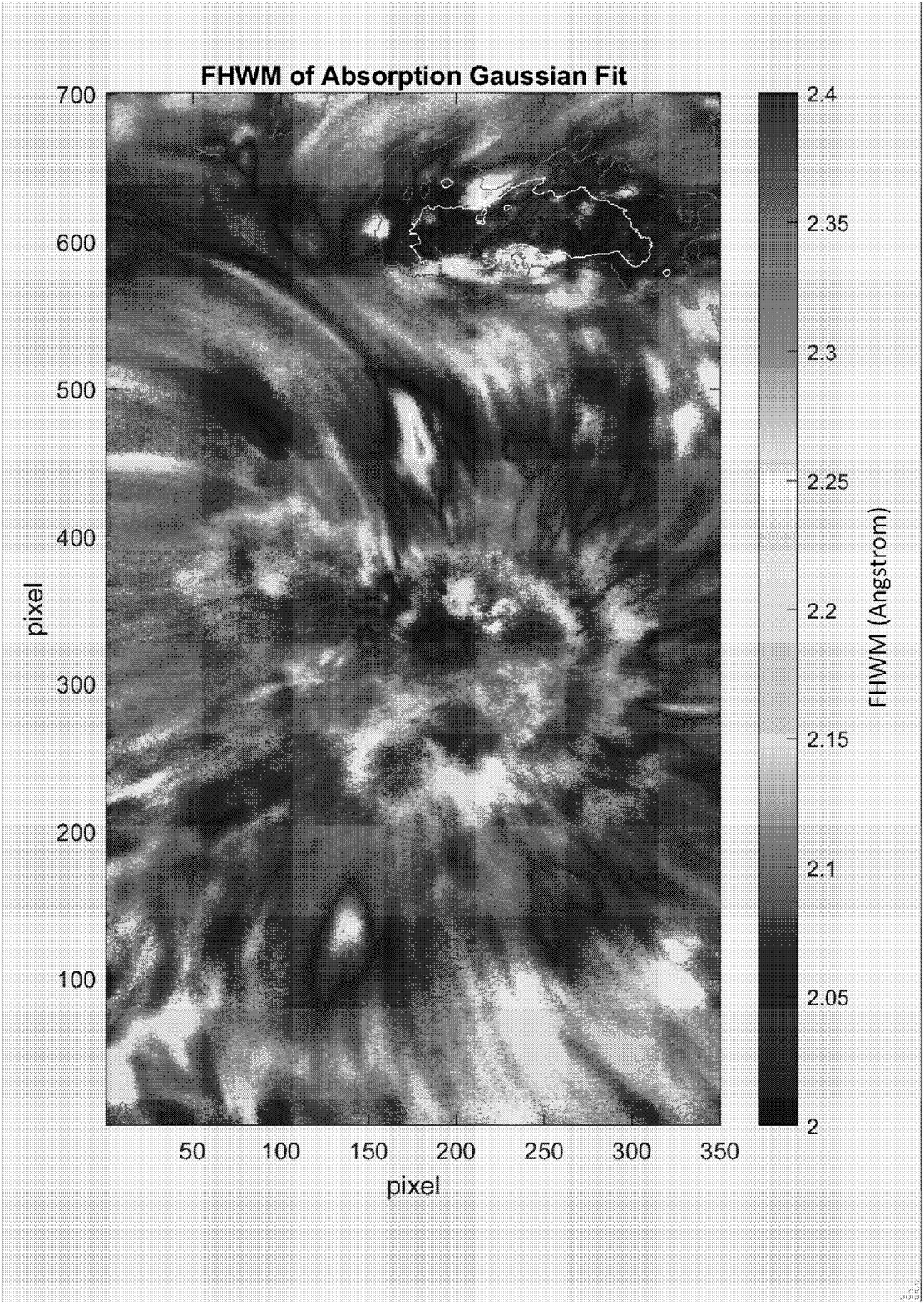}\\
\includegraphics[trim=0 0 0 0, clip, scale=0.16]{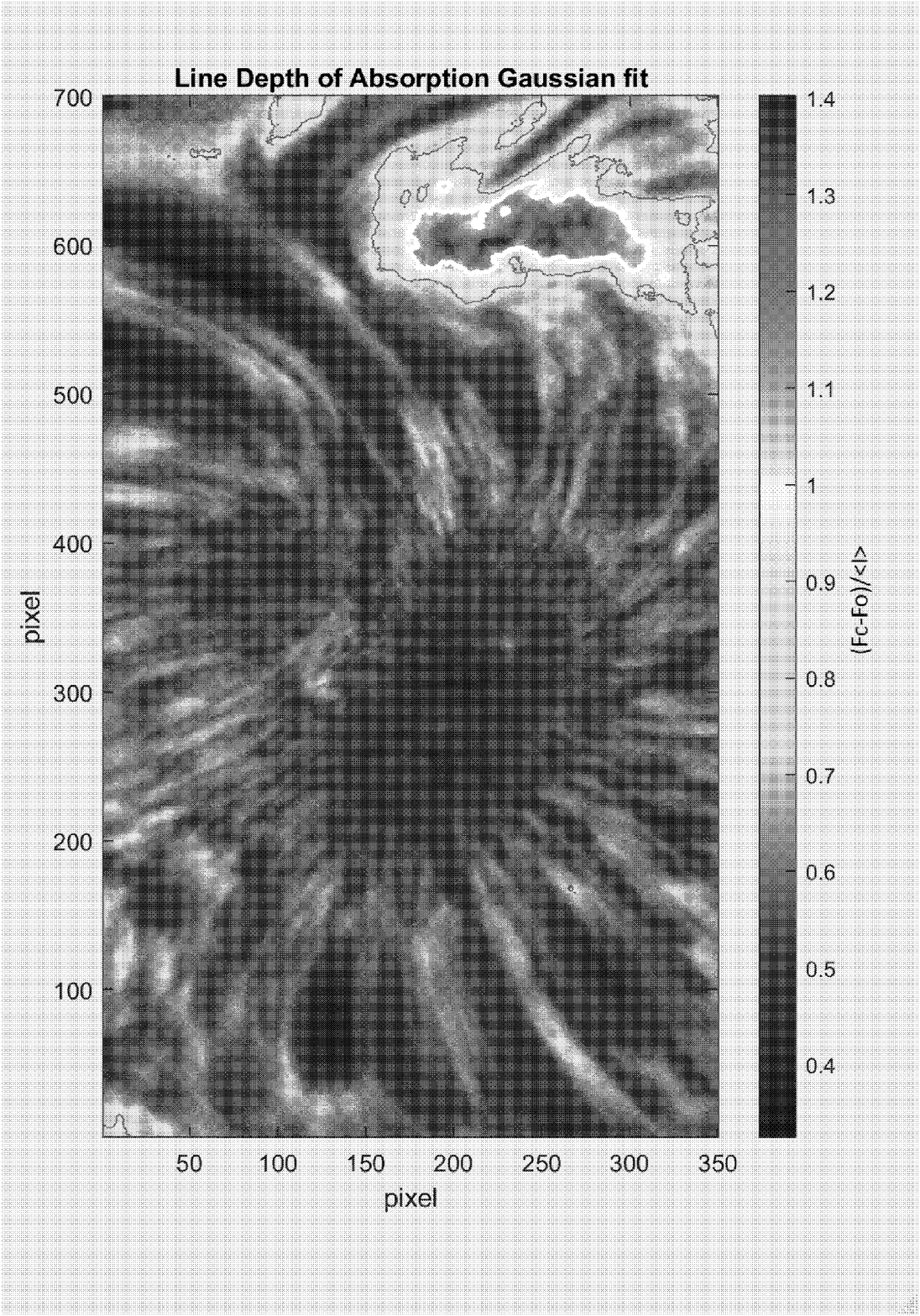}
\includegraphics[trim=0 0 0 0, clip, scale=0.16]{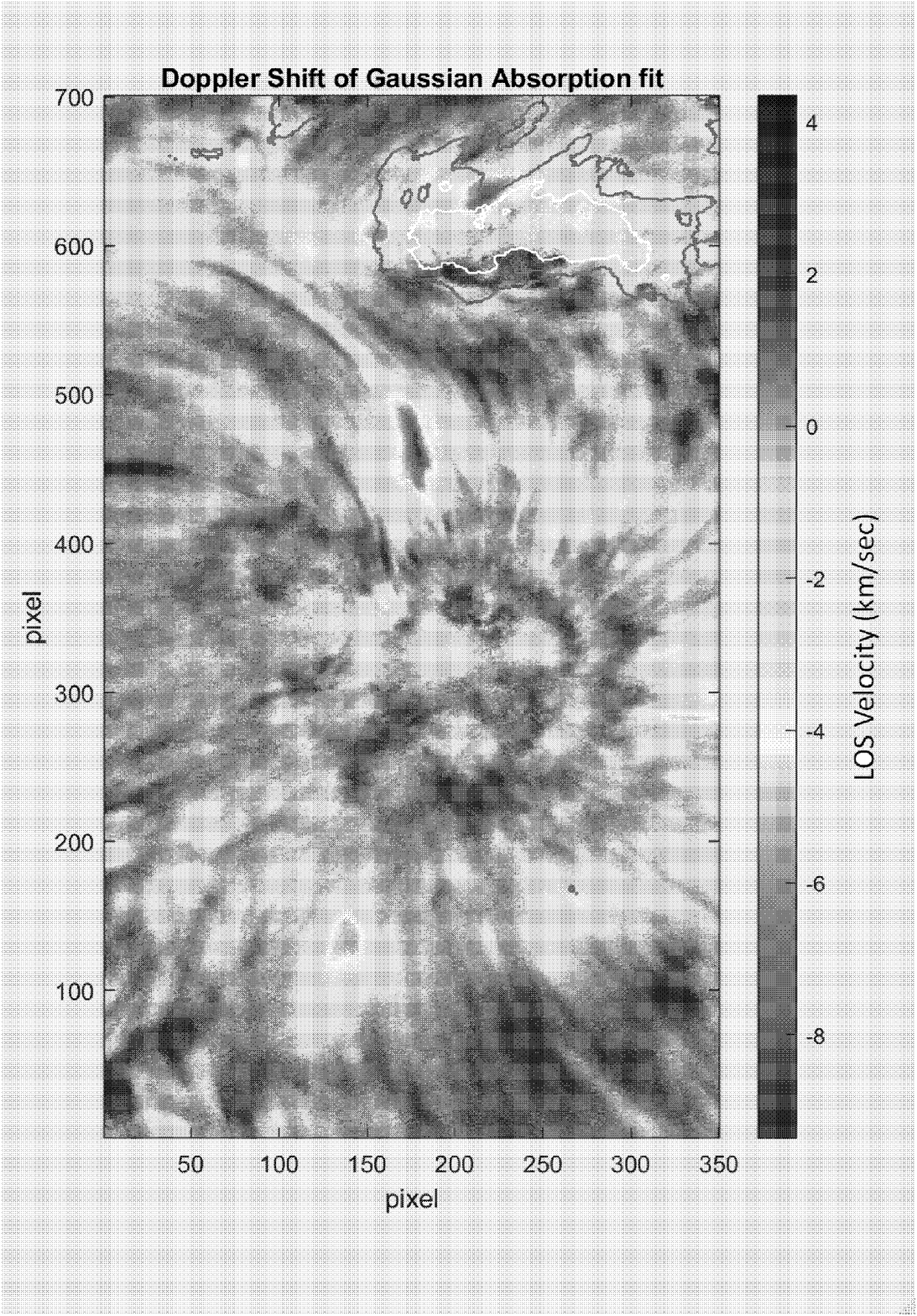}
\caption{Superimposition of the contours of the features corresponding to the facula over the spectral image taken at 656.274 nm.\label{fig7}} 
\end{center}
\end{figure*}

We named each feature by a progressive number from 1 to 16 and we note that they are able to segment in great detail all the structures visible in the FOV at different wavelengths. Moreover, we found an interesting accordance of the features not only with the spectral properties of the segmented photospheric and chromospheric structures, but also with some of their physical parameters. For example, we can see that features n. 1 and 2 identify the chromospheric facula at North. The core of the facula is well segmented by feature n. 1, while feature n. 2 corresponds to the surrounding pattern, visible in the core of the H$\alpha$ line (see the middle panels of Figure \ref{fig1}), where the brightness of the facula is limited by the presence of the neighbouring fibrils (top left panel of Figure \ref{fig7}). In this case, we note that the differences between features n. 1 and n. 2 are not only in their values of intensity in the core of the H$\alpha$ line, but also in the different values of the line depth (bottom left panel of Figure \ref{fig7}). Features n. 1 and n. 2 correspond to LD greater than 0.8 and 1.2, respectively. Instead, we do not see any correlation of these features with LW (top right panel of Figure \ref{fig7} and the LOS velocity (bottom right panel of Figure \ref{fig7}) of the corresponding structures.

\begin{figure*}
\begin{center}
\includegraphics[trim=0 0 0 0, clip, scale=0.16]{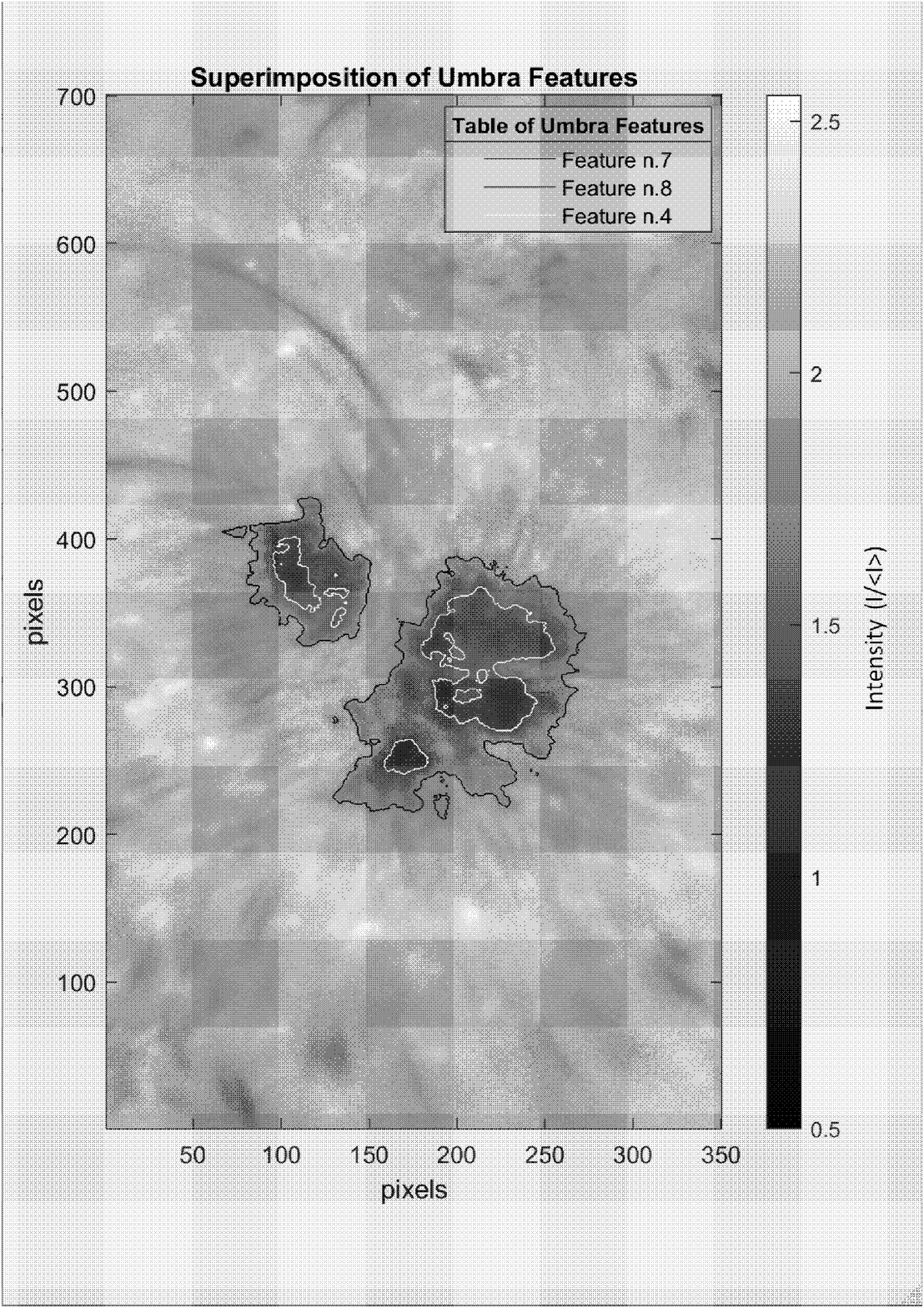}
\includegraphics[trim=0 0 0 0, clip, scale=0.16]{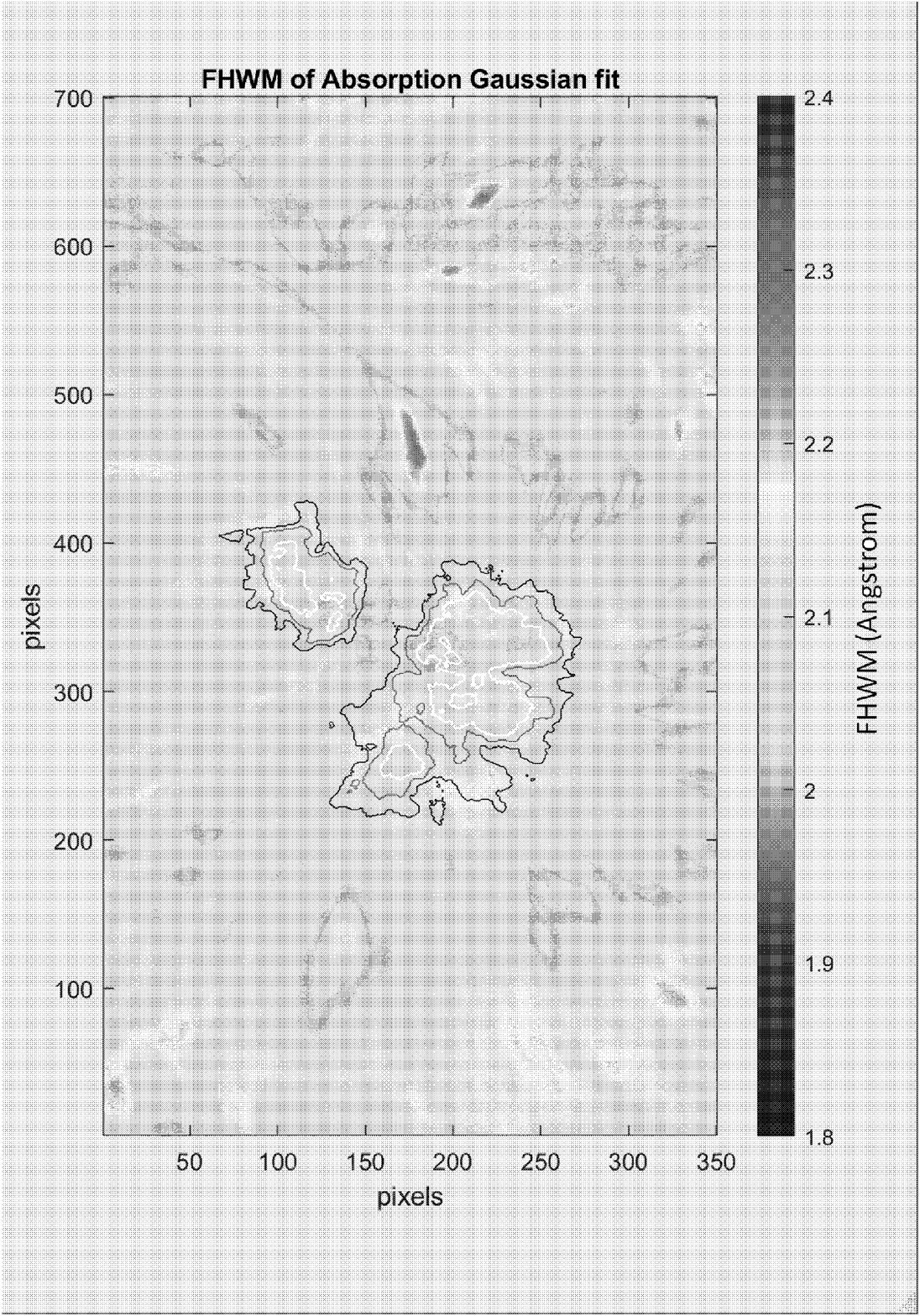}\\
\includegraphics[trim=0 0 0 0, clip, scale=0.16]{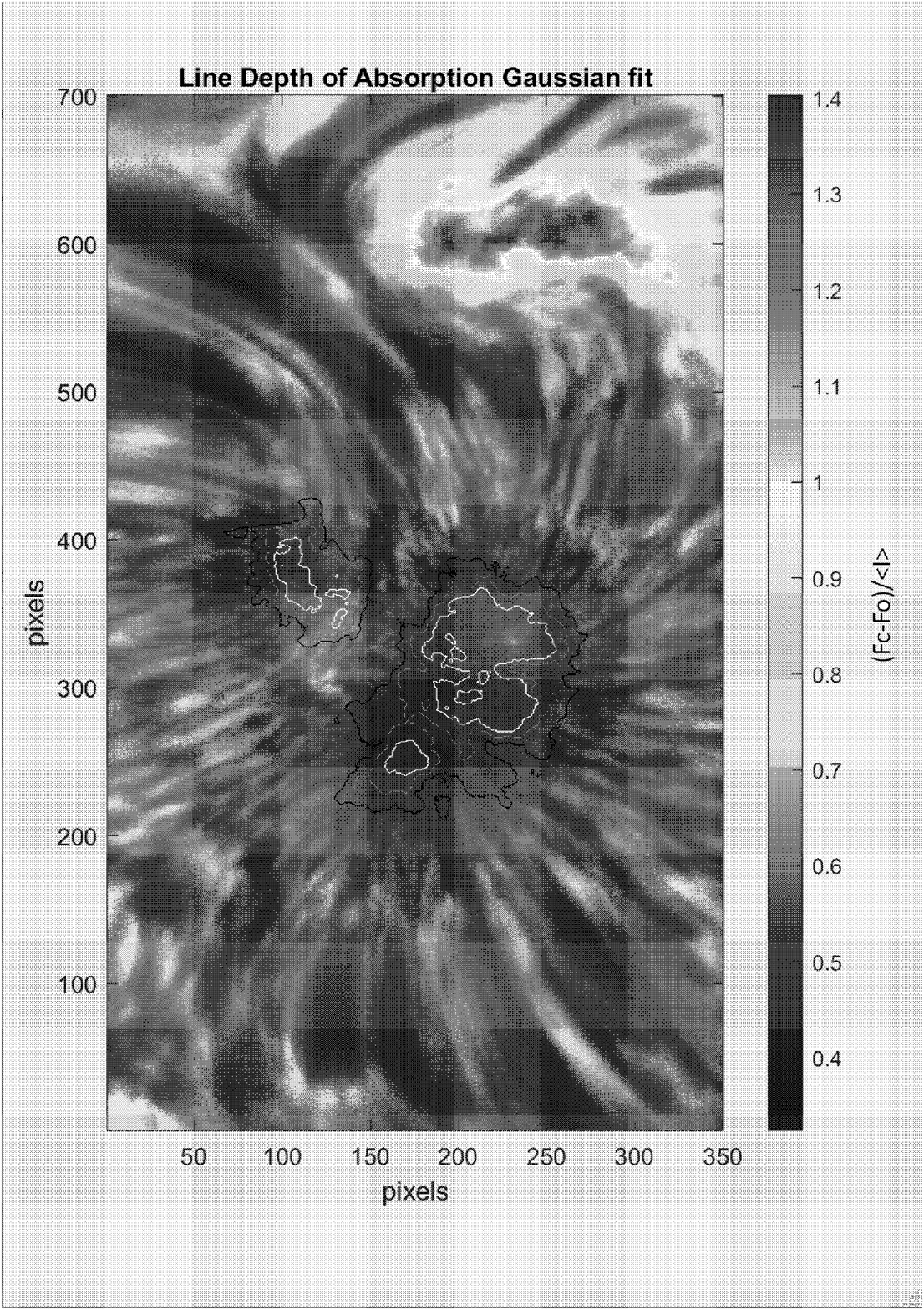}
\includegraphics[trim=0 0 0 0, clip, scale=0.16]{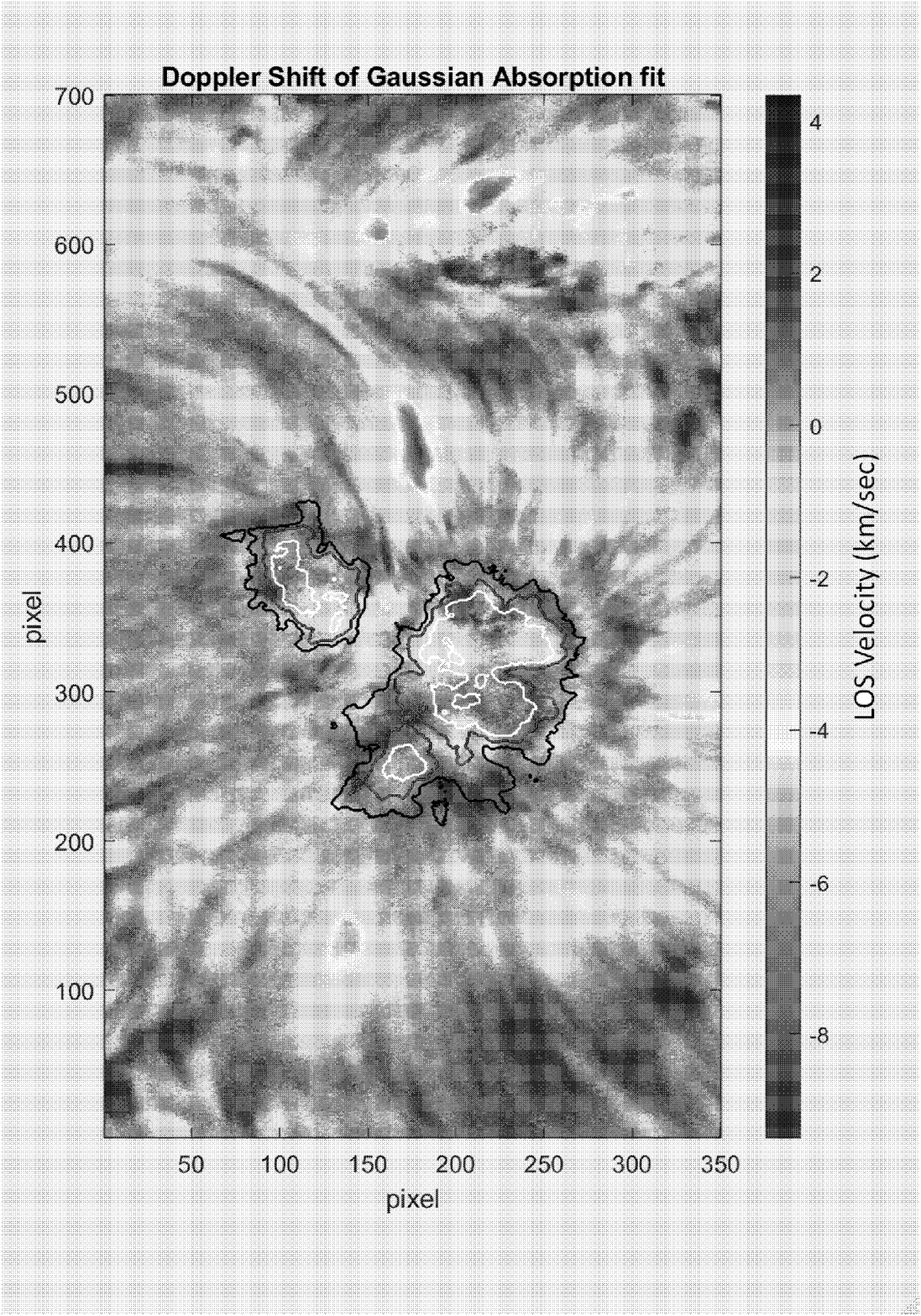}
\caption{Superimposition of the contours of the features corresponding to the sunspot umbra over the spectral image taken at 656.134 nm.\label{fig8}} 
\end{center}
\end{figure*}

The umbra of the sunspot is well segmented by features n. 4, 7 and 8 (see the contours in Figure \ref{fig8}). Combining these three features we are able to cover the whole umbra, while considering only some of them we are able to identify different sub-regions, gradually more extended. Features n. 4 corresponds to the umbra characterized by $I/<I>$ < $1.0$ at 656.134 nm, while features n. 7 and n. 8 segment the portions of the umbra with $1.5$ < $I/<I>$ < $1.7$ and $1.0$ < $I/<I>$ < $1.5$, respectively. LD does not show any significant variation in the umbra (see the bottom left panel of Figure \ref{fig8}). Only LW and the LOS velocity seem to characterize these features. In particular, it is noteworthy that feature n.7 corresponds to the portion of the umbra with a velocity of about 0.5 km s$^{-1}$ (bottom right panel of Figure \ref{fig8}.

\begin{figure*}
\begin{center}
\includegraphics[trim=0 0 0 0, clip, scale=0.16]{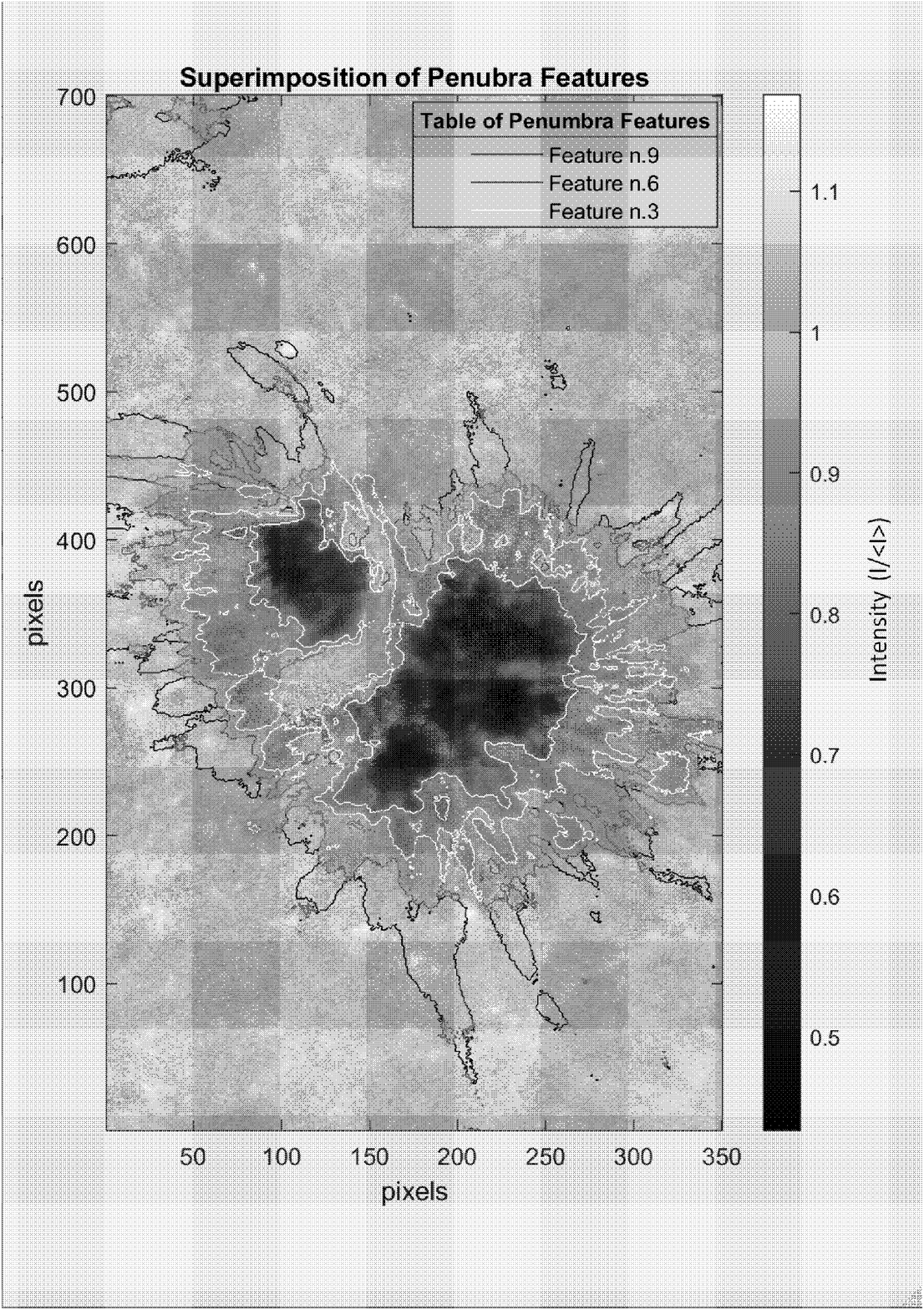}
\includegraphics[trim=0 0 0 0, clip, scale=0.16]{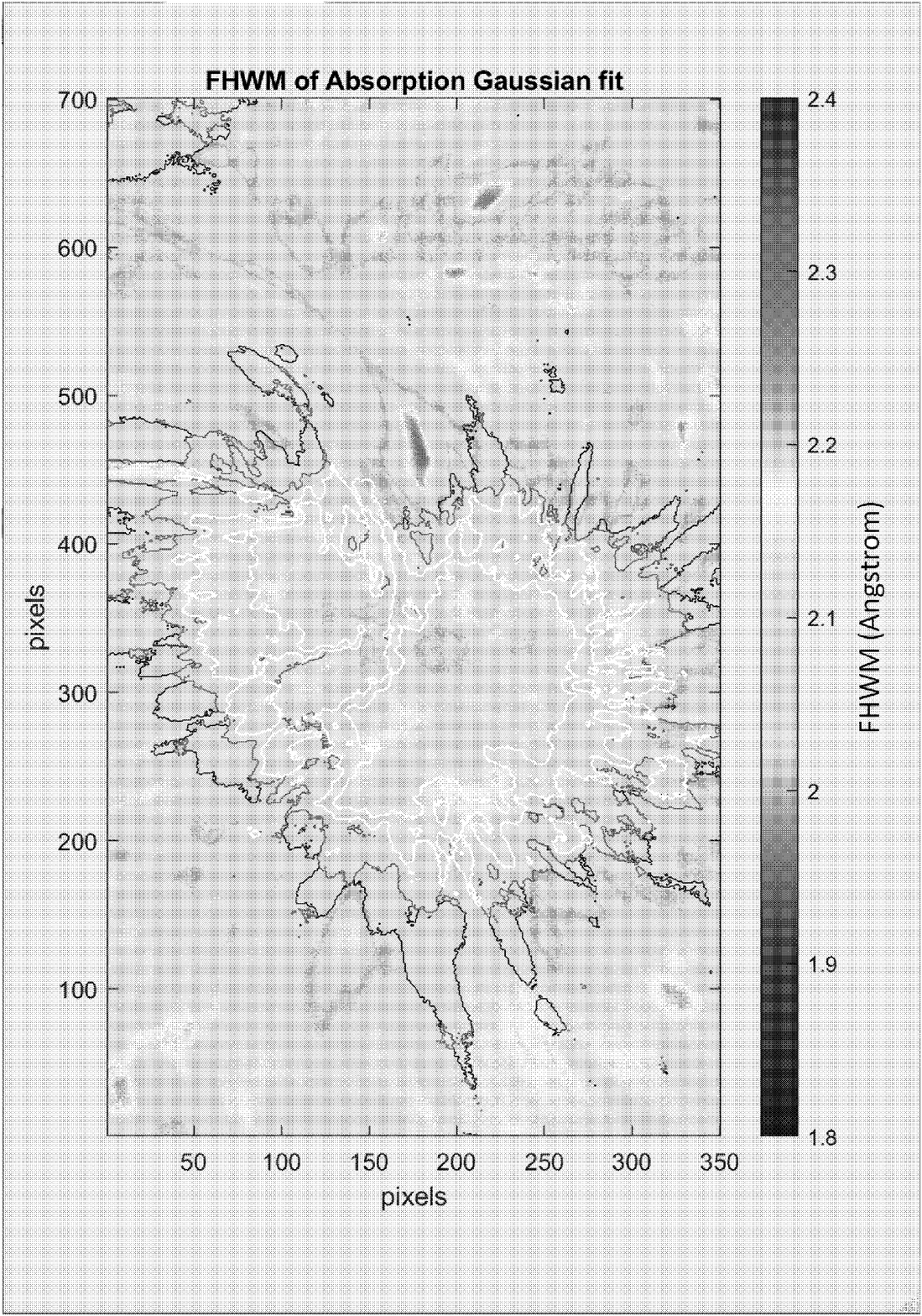}\\
\includegraphics[trim=0 0 0 0, clip, scale=0.16]{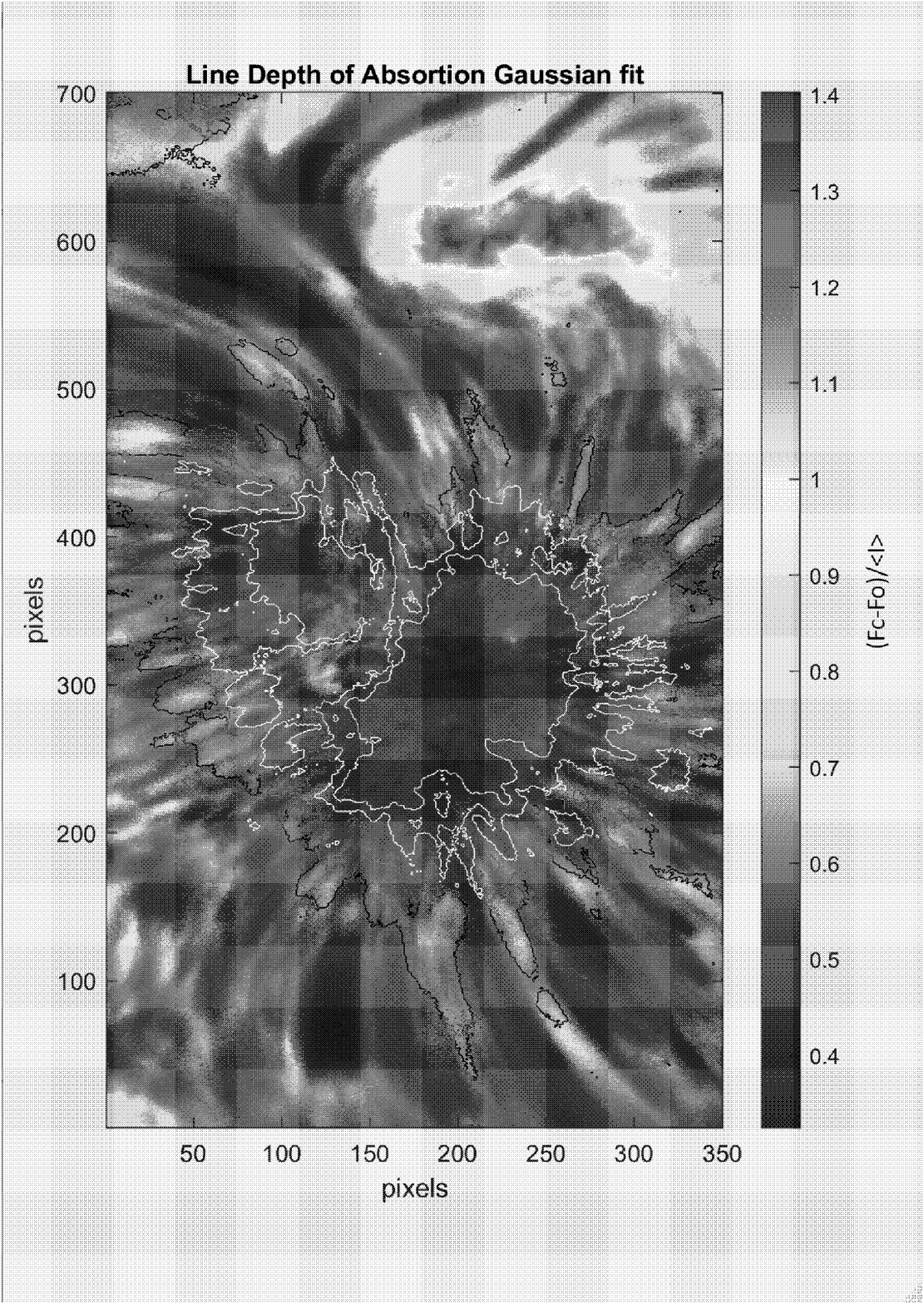}
\includegraphics[trim=0 0 0 0, clip, scale=0.16]{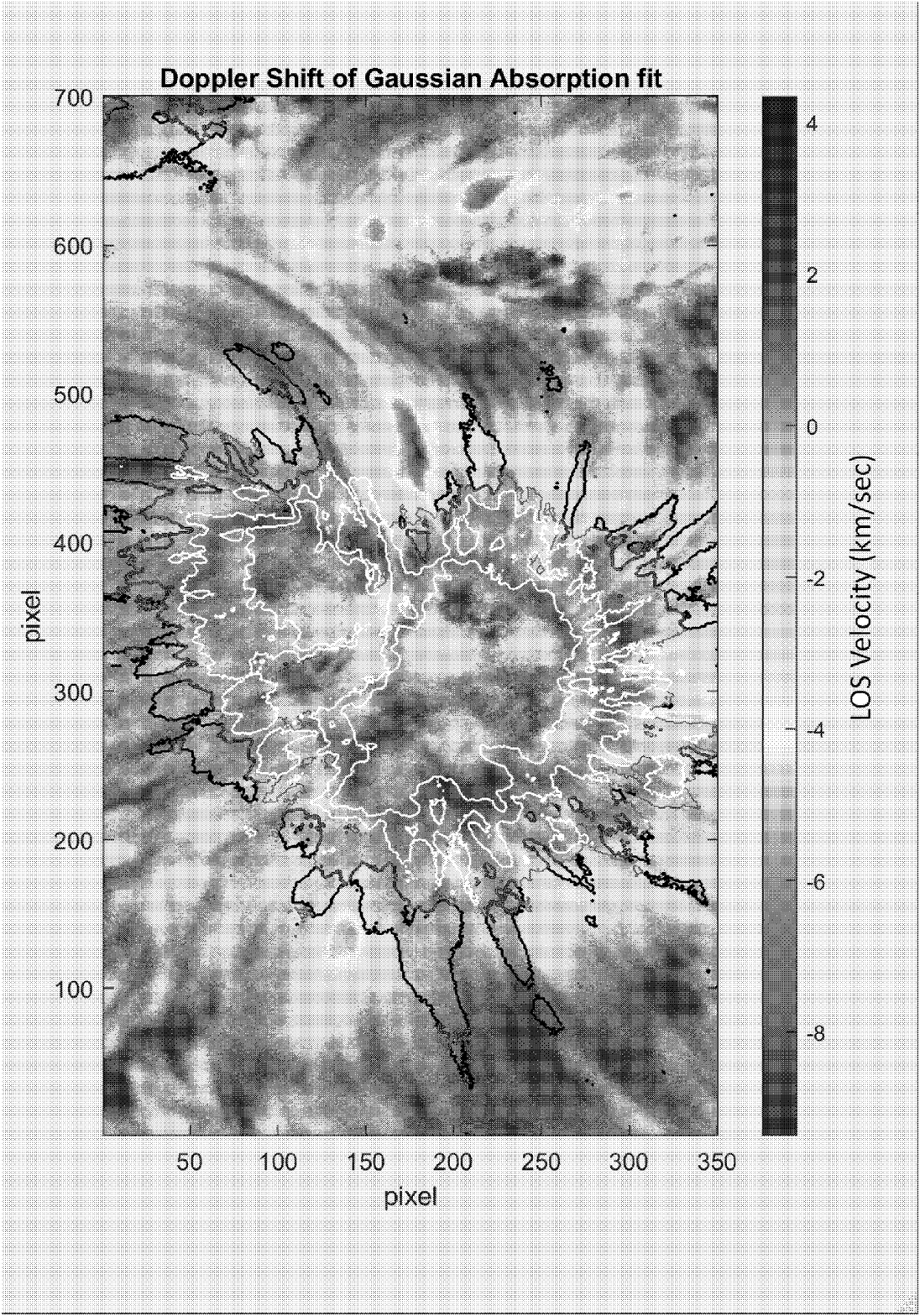}
\caption{Superimposition of the contours of the features corresponding to the sunspot penumbra over the spectral image taken at 656.274 nm.\label{fig9}} 
\end{center}
\end{figure*}

\begin{figure*}
\begin{center}
\includegraphics[trim=0 0 0 0, clip, scale=0.16]{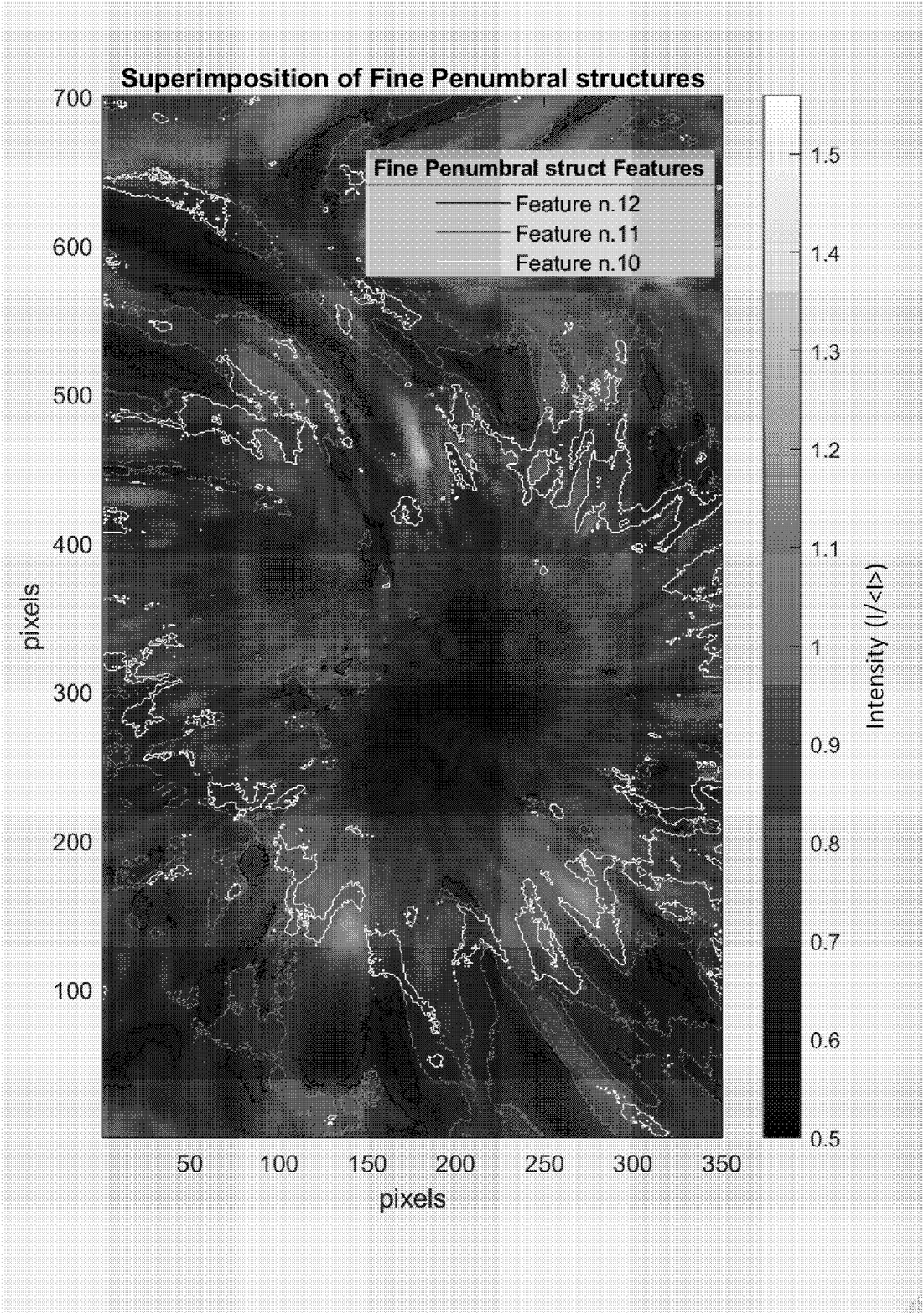}
\includegraphics[trim=0 0 0 0, clip, scale=0.16]{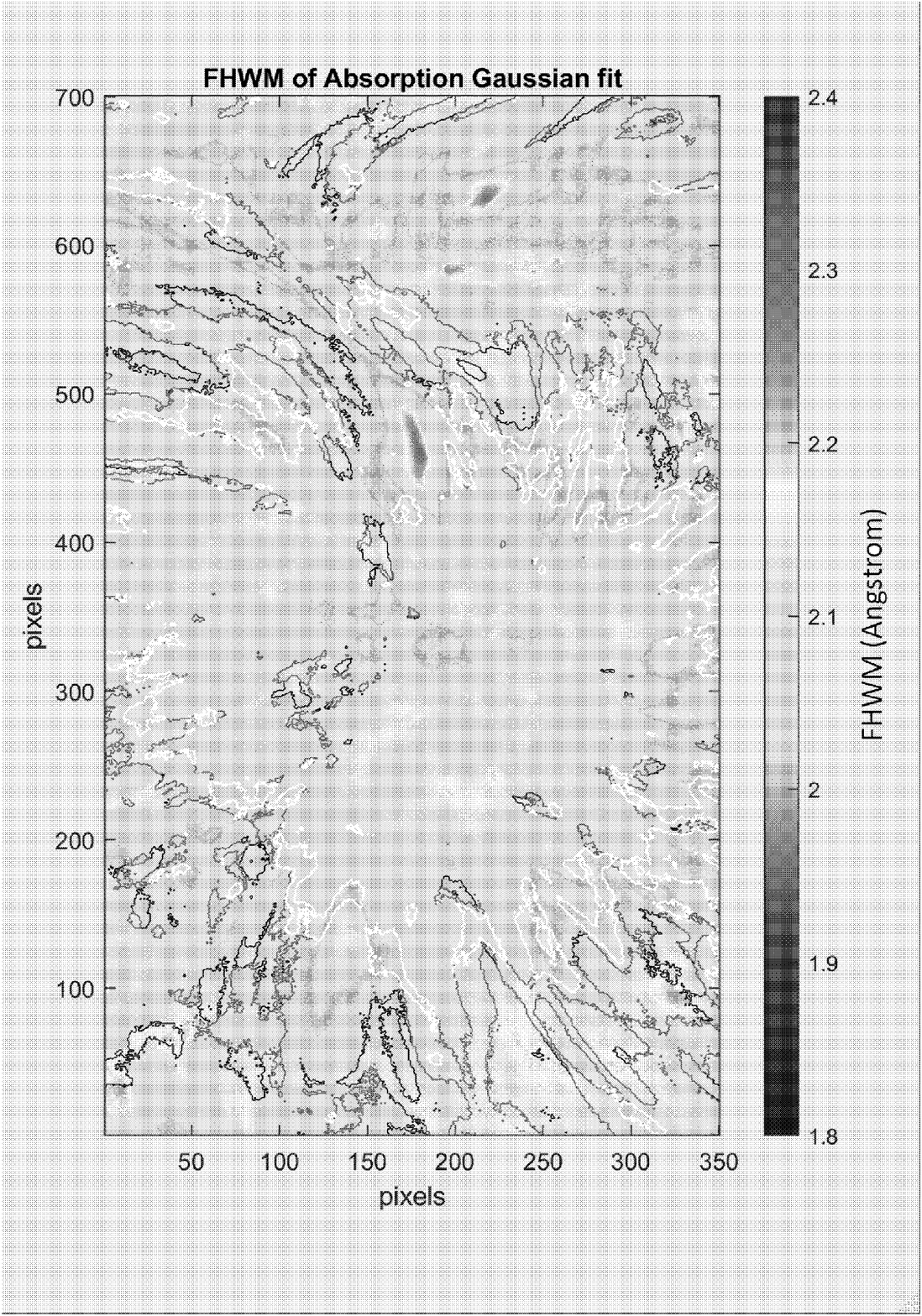}\\
\includegraphics[trim=0 0 0 0, clip, scale=0.16]{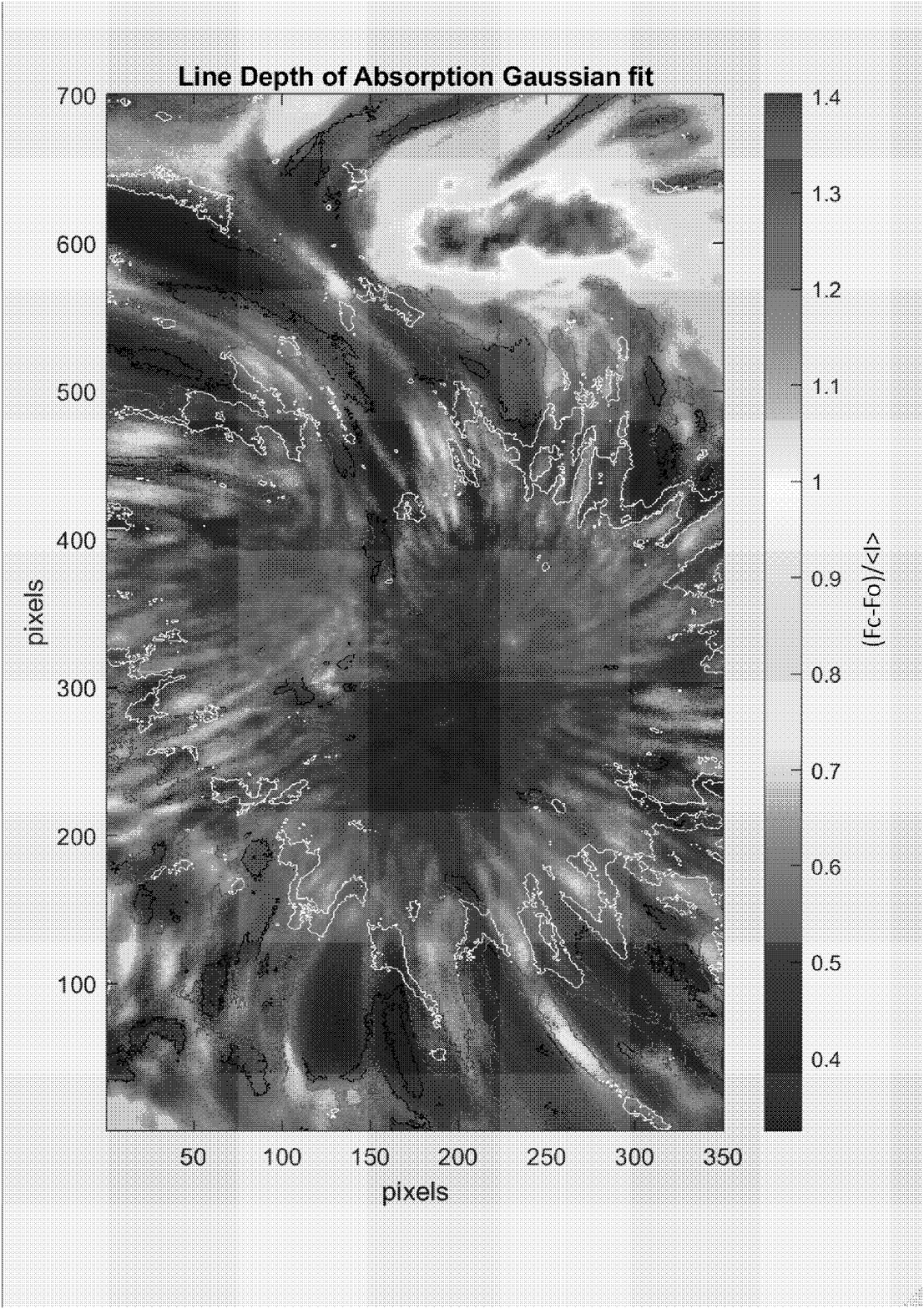}
\includegraphics[trim=0 0 0 0, clip, scale=0.16]{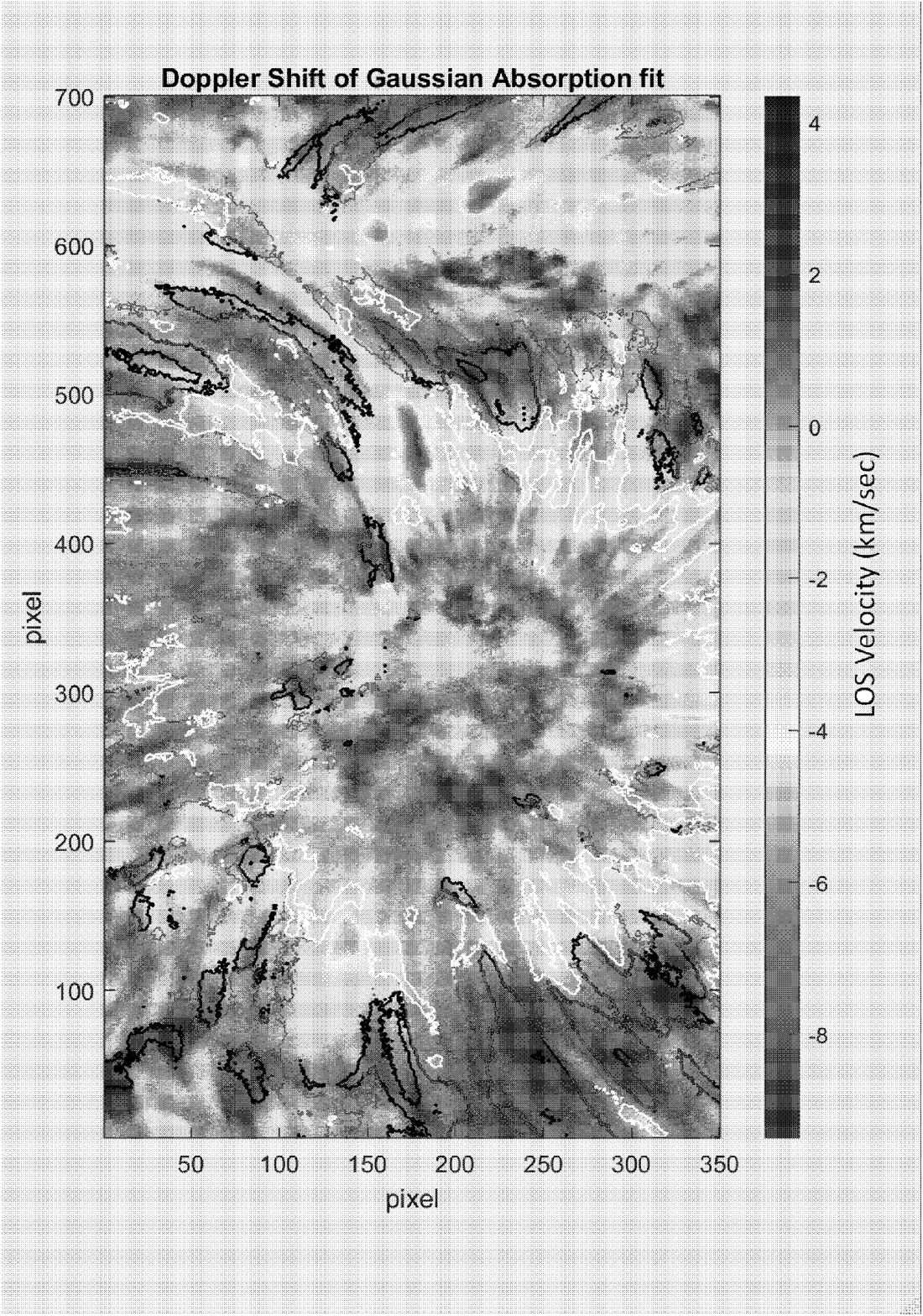}
\caption{Superimposition of the contours of the features corresponding to the superpenumbra over the spectral image taken at 656.274 nm.\label{fig10}} 
\end{center}
\end{figure*}

The penumbra of the sunspot is usually a more complicated structure to be segmented, due to the strong inhomogeneities, in the azimuthal direction as well as along the line-of-sight. It seems that the penumbra consist of dark filaments flanked by lateral brightenings, but also we can see bright filaments formed by a few penumbral grains that are radially aligned. Nevertheless, features n. 3, 6, 9, 10, 11 and 12 are useful to identify the different parts of the penumbra. In particular, features n. 9 and 10 seem to be useful to discriminate among dark and bright regions forming the fine structure of the penumbra at photospheric level, while feature n. 11 and 12 identify the superpenumbra at chomospheric level (see Figure \ref{fig9}). Taking into account the physical properties corresponding to these features, we can see that the features n. 3, 6 ad 9 correspond to different levels of LW, LD and LOS velocity, although due to the Evershed flow \citep{Ever09} we cannot associate specific values of LW and DS to each feature. Only the LD shows different levels for each feature, in fact features n. 3, 6  and 9 are characterized by LD around 0.4, 0.5 and 0.7, respectively. For the superpenumbra we can see also that features n. 10, 11 and 12 correspond to different level of LW (the top right panel of Figure \ref{fig10}), LD (bottom right panel of Figure \ref{fig10}) and LOS velocity (bottom left panel of Figure \ref{fig10}).

\begin{figure*}
\begin{center}
\includegraphics[trim=0 0 0 0, clip, scale=0.16]{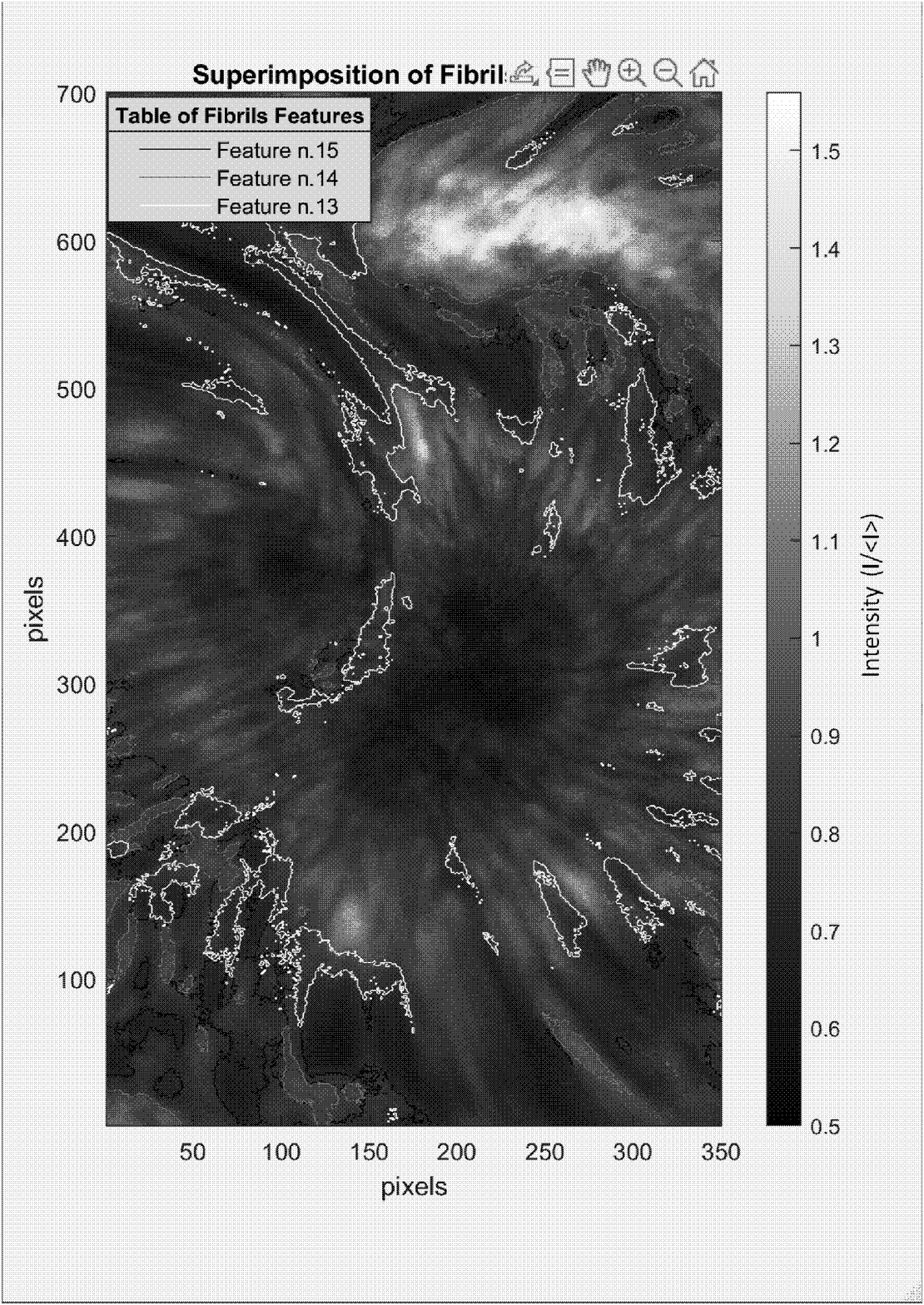}
\includegraphics[trim=0 0 0 0, clip, scale=0.16]{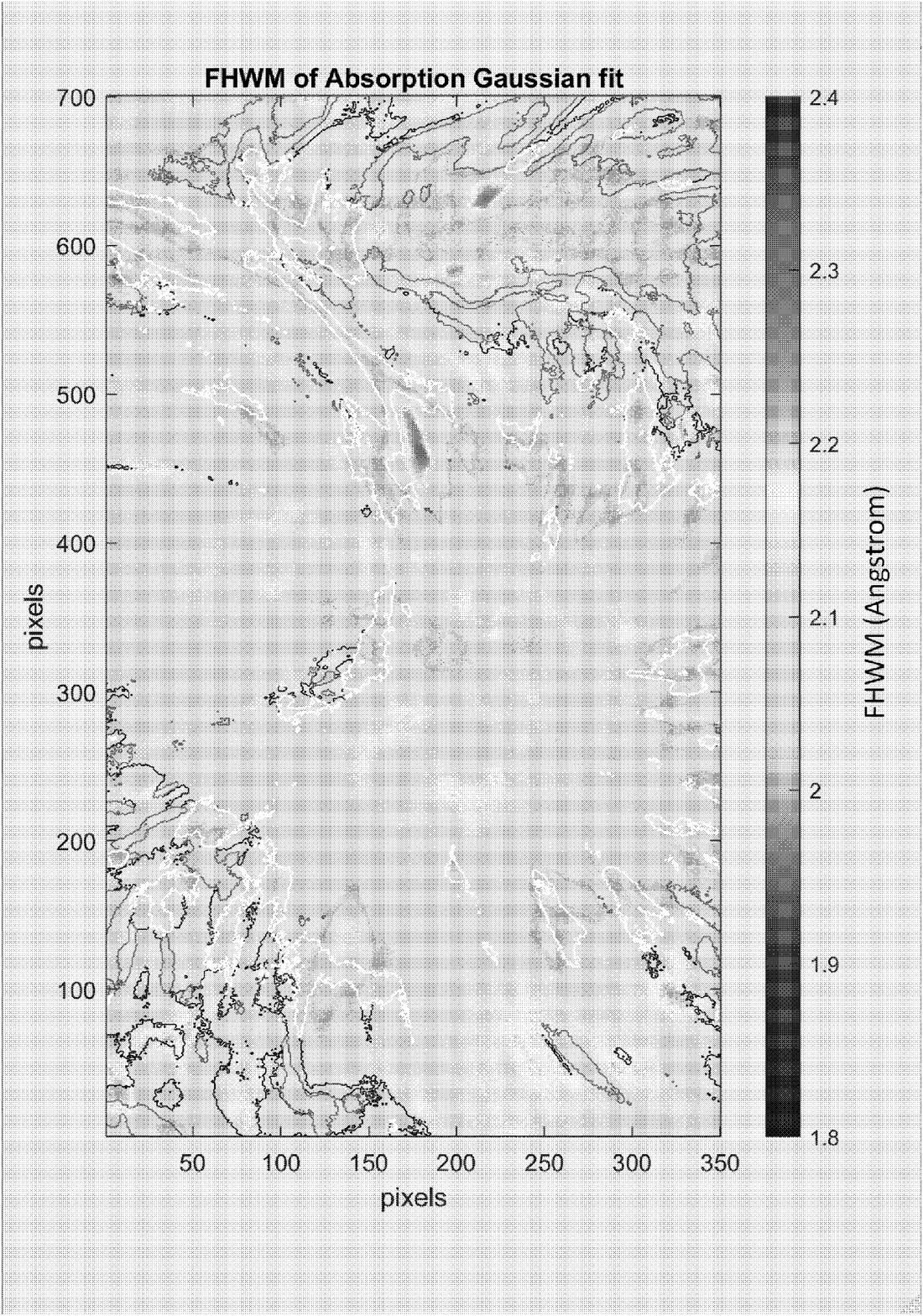}\\
\includegraphics[trim=0 0 0 0, clip, scale=0.16]{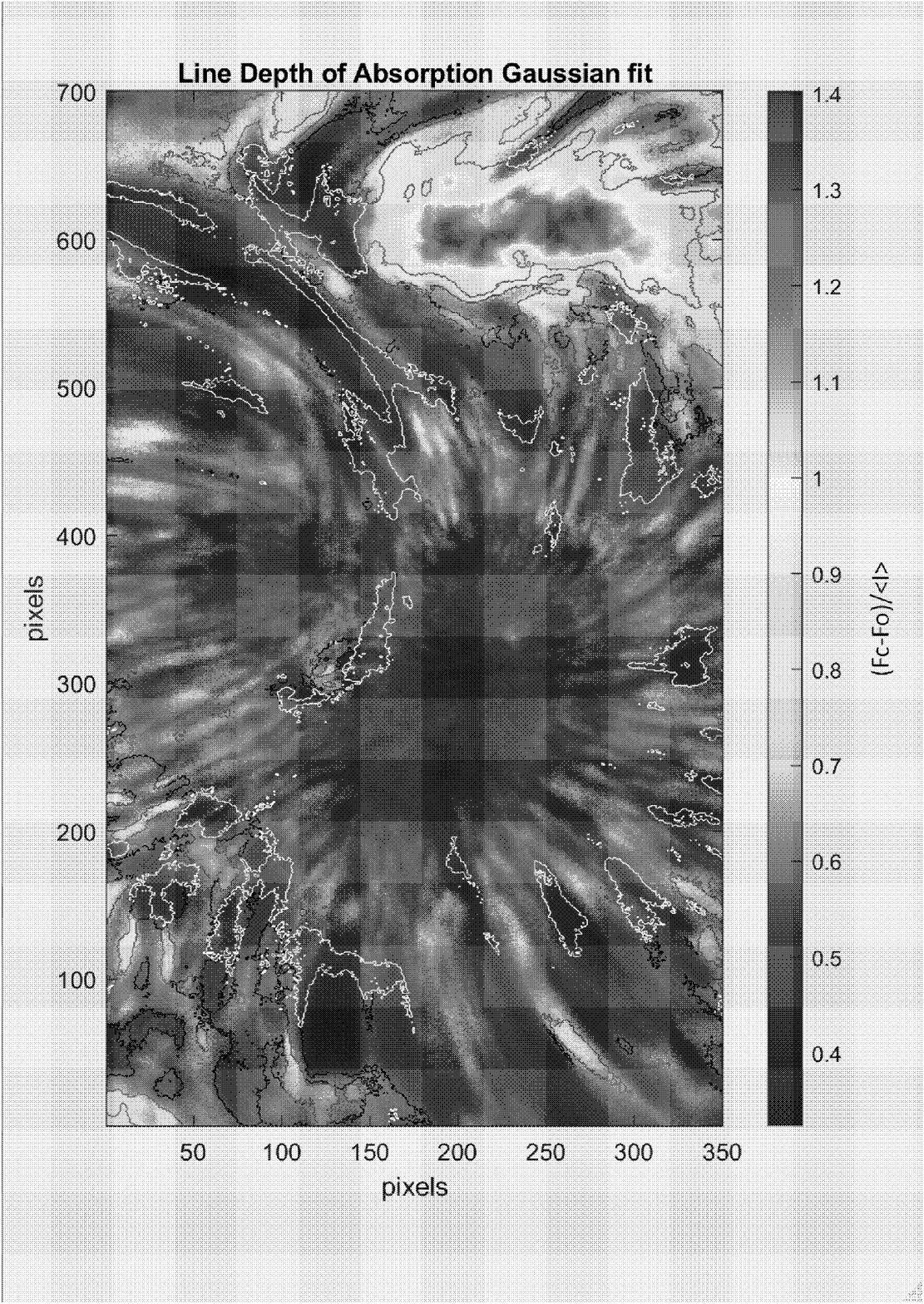}
\includegraphics[trim=0 0 0 0, clip, scale=0.16]{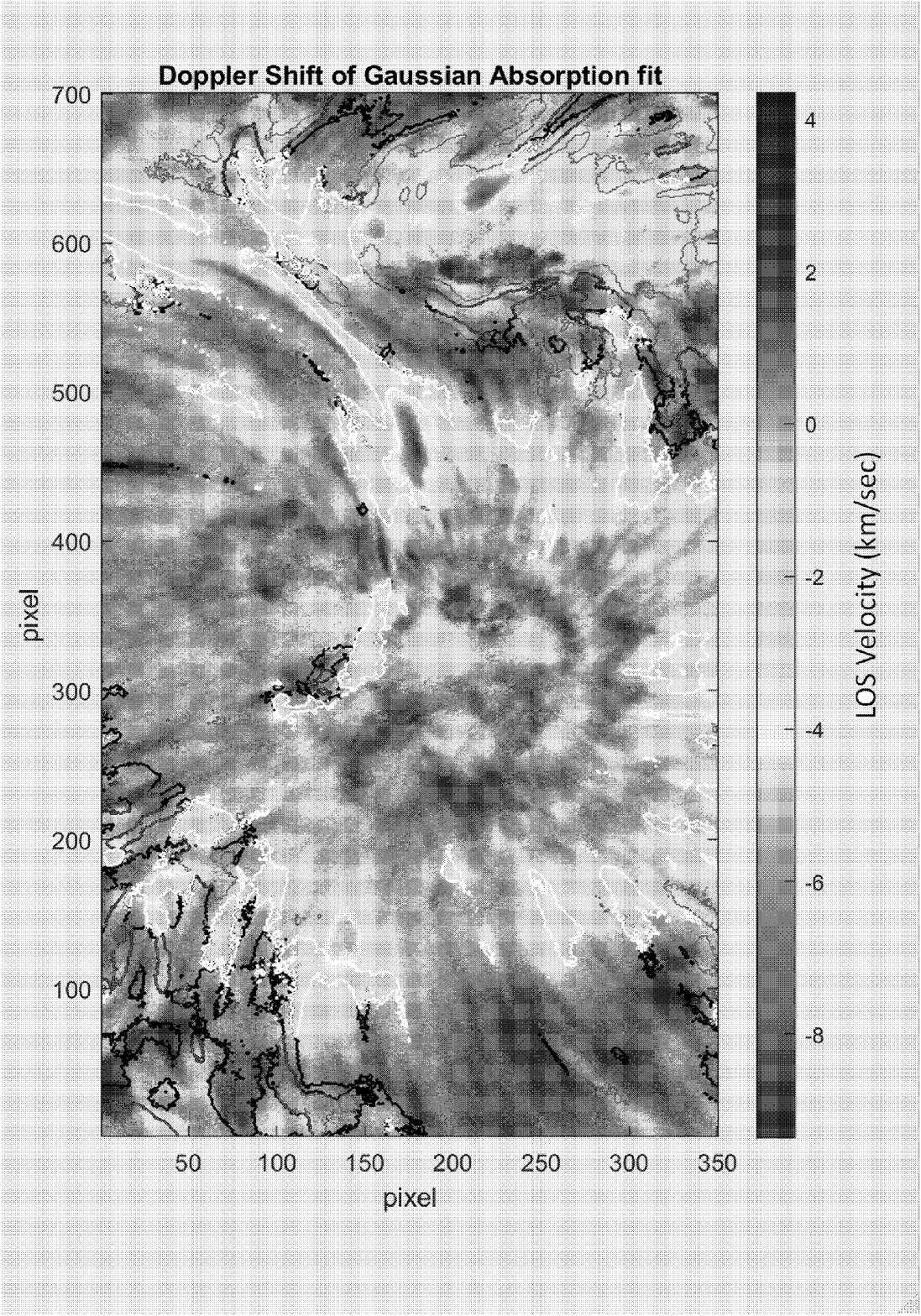}
\caption{Superimposition of the contours of the features corresponding to the fibrils over the spectral image taken at 656.274 nm.\label{fig11}} 
\end{center}
\end{figure*}

Features n. 13, 14 and 15 correspond to the fibrils observed in the surrounding region around the sunspot, as depicted in Figure \ref{fig11}. In particular, feature n. 13 is characterized by a FWHM of the H$\alpha$ line of about 2.0 \AA, an LD of about 0.4 and a positive LOS velocity (around 1 km s$^{-1}$). Instead, features n. 14 correspond to the portion of the fibrils with LW $\sim$ 2.1 \AA{} and LD $\sim$ 0.7. Positive values of the LOS velocity characterize features n. 15 (see bottom right panel of Figure \ref{fig11}).

\begin{figure*}
\begin{center}
\includegraphics[trim=0 0 0 0, clip, scale=0.16]{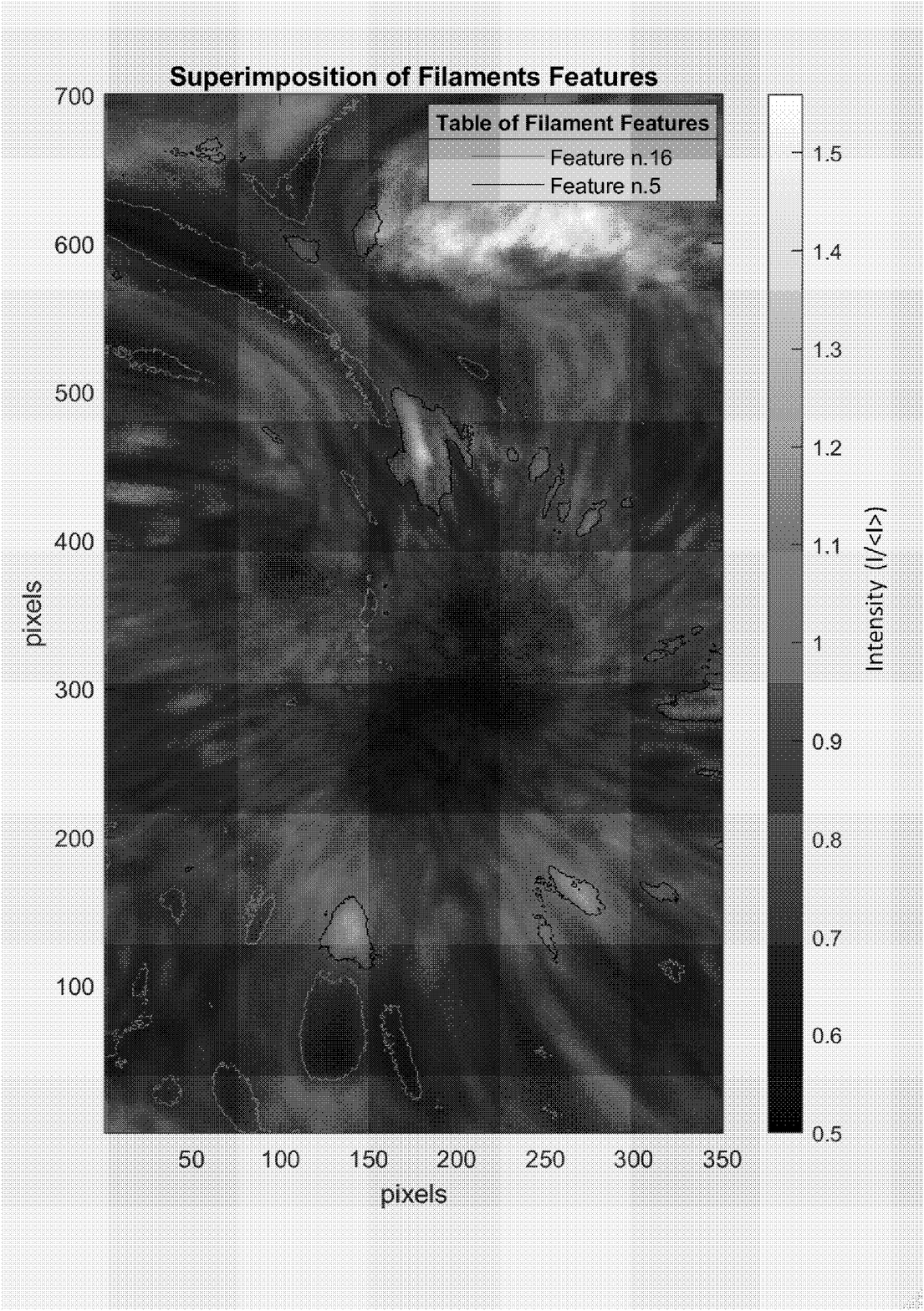}
\includegraphics[trim=0 0 0 0, clip, scale=0.16]{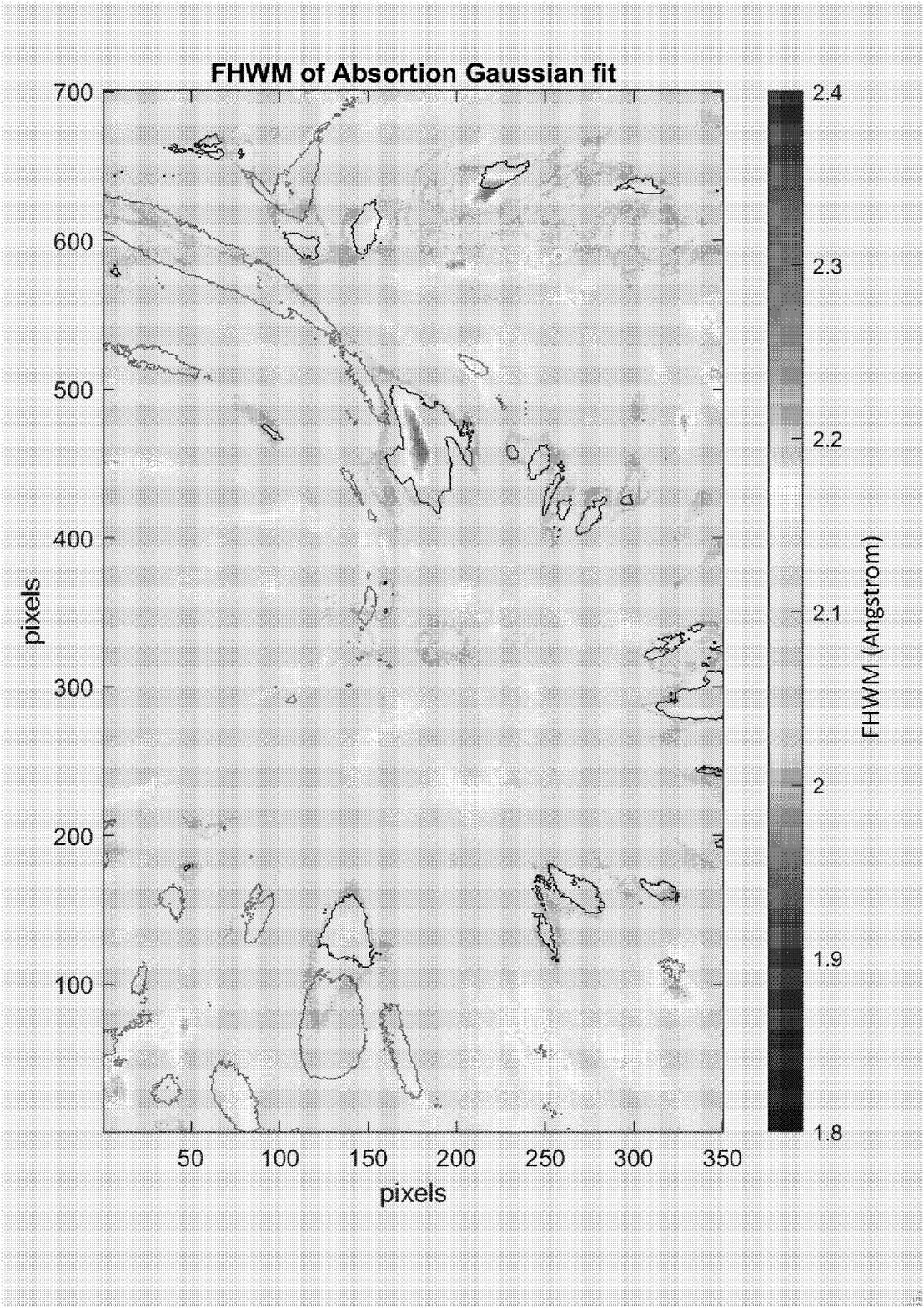}\\
\includegraphics[trim=0 0 0 0, clip, scale=0.16]{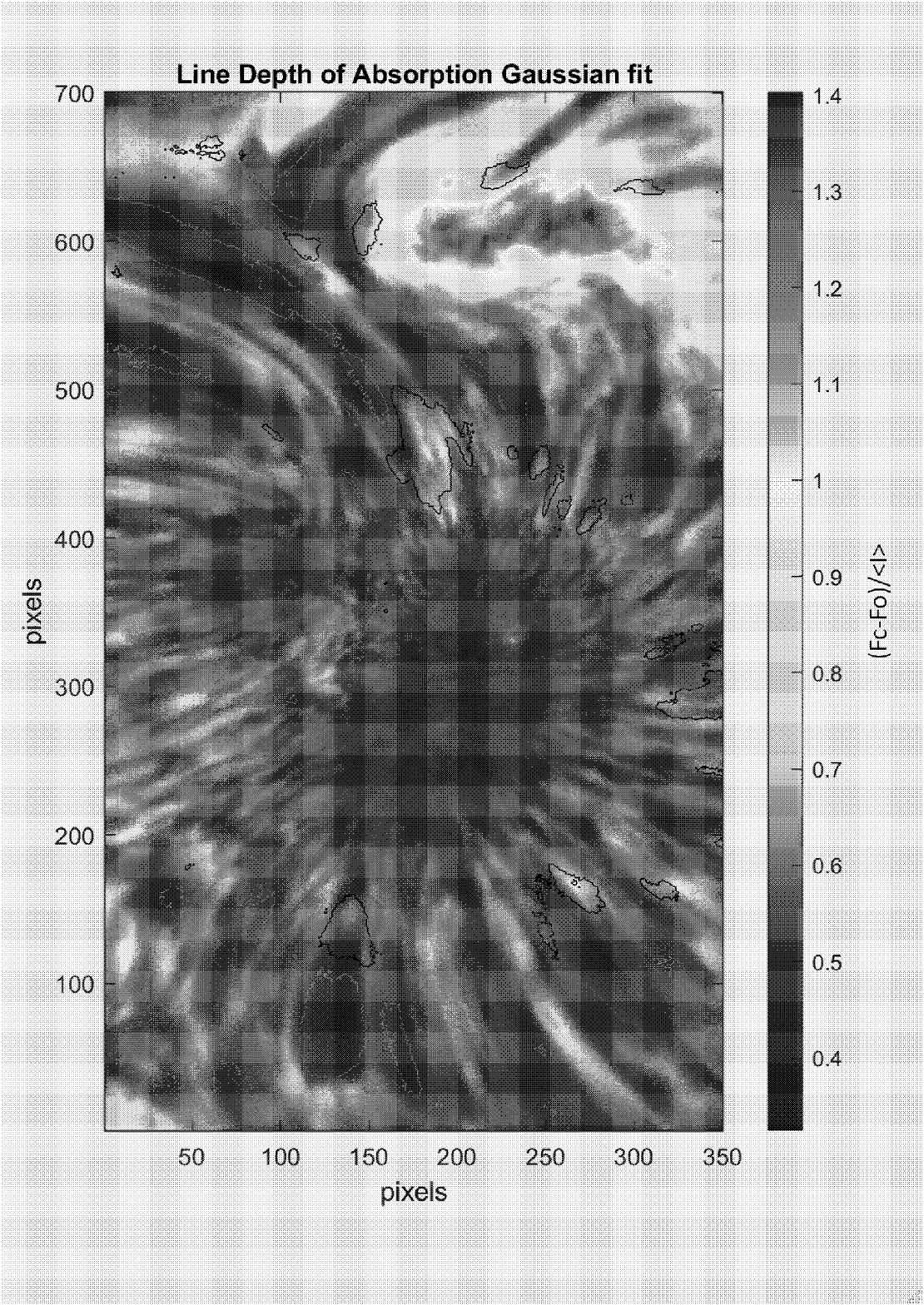}
\includegraphics[trim=0 0 0 0, clip, scale=0.16]{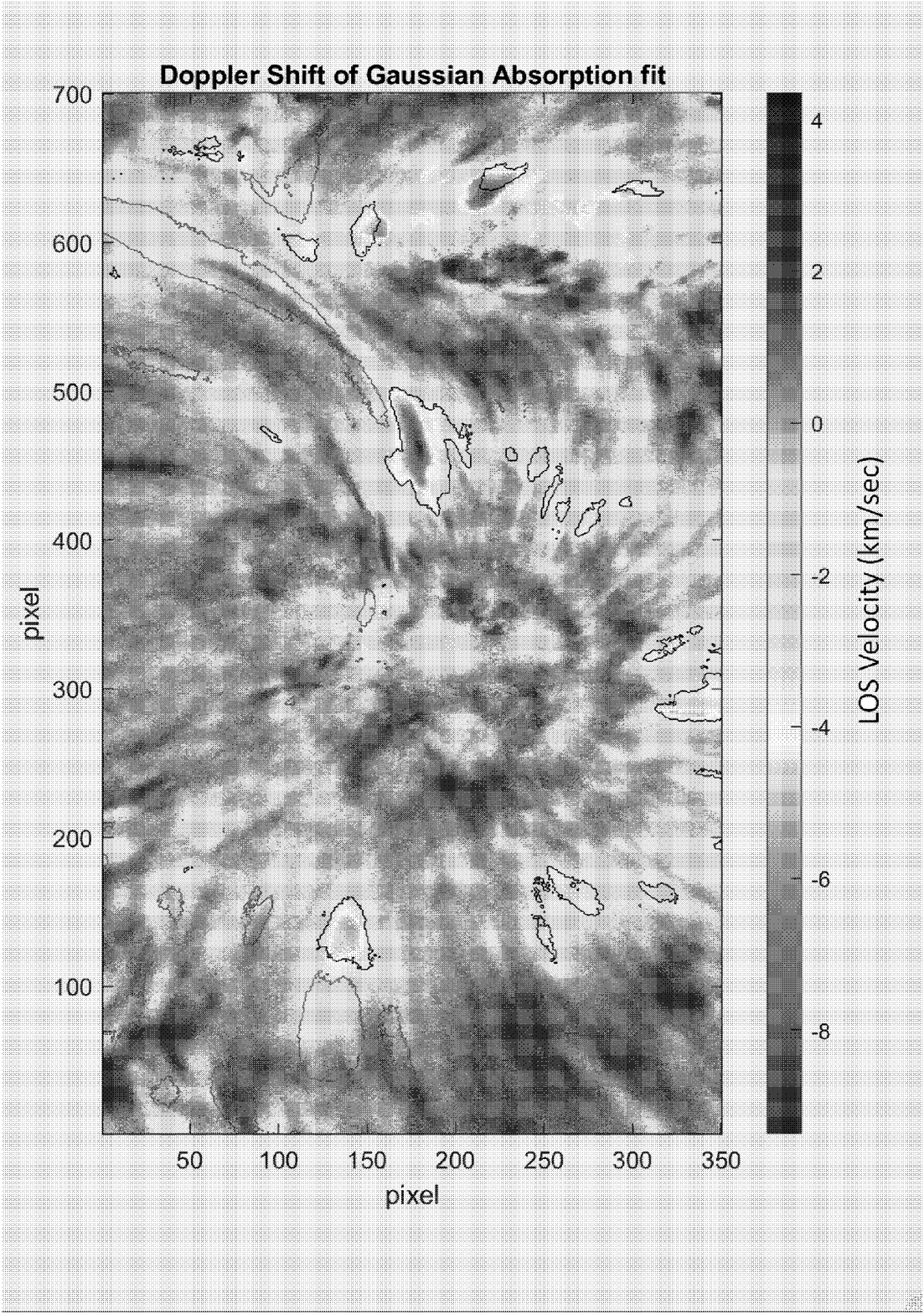}
\caption{Superimposition of the contours of the features corresponding to the filaments over the spectral image taken at 656.274 nm..\label{fig12}} 
\end{center}
\end{figure*}

Finally, feature n.16 can be considered to study the filament portions visible in the FOV. In fact, we clearly recognize the elongated shape of a filament in the top left corner of the bottom right panel of Figure \ref{fig12}. Instead, feature n. 5 seems to identify the brighter regions corresponding to the footpoints of the filaments, the so called umbral filaments, i.e., structures where the increase of the plasma emission can be due to the accumulation of plasma coming from higher solar atmospheric levels into the photosphere and the low chromosphere \citep[see][for more details]{Gug17, Gug19}. This is confirmed by the LOS velocity map (bottom right panel of Figure \ref{fig12}), where we see that the black contours (feature n. 5) correspond to the stronger red shifted areas with velocity lower than -8 km s$^{-1}$, while the red contours (feature n. 16), indicating the upper parts of the filaments, are characterized by LOS velocities closer to 0 km s$^{-1}$.

\section{Conclusions}

The use of Machine Learning methods becomes increasingly important in the world of big data, for the ability to process data in a massive and intelligent way by increasingly performing processors. 

In this work, we applied for the first time the SOM method to high resolution images of the solar atmosphere. We exploited the main advantage of a bi-dimensional spectrometer based on Febry-Perot devices, such as IBIS. Using high spectral and spatial resolution images taken at different wavelengths along the same spectral line (in our case the H$\alpha$ line), we were able to identify different structures of the solar atmosphere that depend on the plasma and magnetic field properties. The application of the SOM technique was perfectly suitable to segment the different images and to isolate among these the structures for a subsequent analysis. We remark that this method seems to obtain also a good correspondence between the features and the physical properties of the identified structures. In fact, we found some differences in terms of physical or spectral properties (e.g. line core depth, line width and doppler shift) exhibited by features segmenting different portions of the same solar structure.

The advantages of the algorithm described in this paper are certainly manifold. First of all the developed algorithm is very efficient: using our dataset, formed by 17 images of 350 $\times$ 700 pixels, its computation  requires no pre-elaboration of data, took less than one minute for a SOM calculation with 200 epochs, using an octa-core i9 processor with 2.3 GHz clock. Moreover the algorithm can be parallelized and can be exported on heterogeneous calculation platforms having greater computational efficiency (e.g., GPU, FPGA) than the platform used to find the prototype results achieved.
In addition, this algorithm can be implemented on front-end devices that allow a complete automatic analysis of the data collected and its processing in an almost real-time regime {\bf\citep[]{Hug19}}, so monitoring of the regions of interest could be automatic and independent of the operator.
In comparison to other methods reported in literature, we note that the weakness of this method concerns the different position of a feature into the SOM lattice, mainly after the elaboration of subsequent datasets. This issue has some consequences when the user needs to compare or elaborate features in different frames in order to perform their tracking or to measure the evolution of their physical properties. For this reason, other segmentation algorithms, like Yet Another Feature Tracking Algorithm \citep[YAFTA,][]{Welsch04}, provided certainly better performances for this scope. However, research activity in order to identify some topics of the feature and run a next cascade supervised and fast network is in progress.

 Moreover we believe that the segmentation carried out is very accurate and is based on the determination of possible clusters emerging from the data itself. Unlike other methods \citep{Aschw10, DeV15, Schad17, Per13, Zar05} the segmentation reveals all the possible structures in one run and it is verified on the basis of objective parameters that make the segmented regions complementary and physically well connoted, for this reason it is known as semantic segmentation.

This work represents a first step towards the application of the SOM technique to astrophysical data sets. Taking into account that other skeletonization methods are not able to identify so many structures at the same time and that such a result could not be achieved by a simple selection of several thresholds, we think that the SOM technique may find very useful applications to the analysis of this kind of data in the context of the space weather field. In fact, we also plan to apply the same technique to other solar data. We think that its application to full disk images taken at different wavelengths corresponding to different layers of the solar atmosphere, e.g.,  images taken by the Solar Dynamics Observatory \citep[SDO;][]{Pes12}, may reveal further potentialities for this kind of unsupervised approach \citep[]{Verb13}.

Finally, the possibility to follow the evolution of those features identified by the SOM technique over several days of observation could be a useful tool to find new parameters which are able to forecast the occurrence of eruptive events in solar active regions. 

\section*{Acknowledgements}
The authors wish to thank the DST staff for its support during the observing campaigns. The research leading to these results has received funding from the European Union’s Horizon 2020 research and innovation programme under grant agreement no. 824135 (SOLARNET project). This work was supported by the Italian MIUR-PRIN 2017 on Space Weather: impact on circumterrestrial environment of solar activity and by the Space Weather Italian COmmunity (SWICO) Research Program.
A special thanks to MOSAICo projects (Metodologie Open Source per l'Automazione Industriale e delle procedure di CalcOlo in astofisica) funded by Italian MISE, Ministero Sviluppo Economico).

\section*{Data Availability}

The data set used in this Paper is available at http://ibis.oa-roma.inaf.it/IBISA/ , in the archive containing data acquired with IBIS during some observing campaigns. The archive IBIS-A has been realised in the framework of the FP7 SOLARNET3 High-resolution Solar Physics Network that aims at integrating the major European infrastructures in the field of high-resolution solar physics, as a step towards the realisation of the European Solar Telescope \citep{Col10}.



\bibliographystyle{mnras}








\bsp	
\label{lastpage}
\end{document}